\documentclass[preprints,article,accept,moreauthors,pdftex,10pt,a4paper]{mdpi}

\usepackage[utf8]{inputenc}
\usepackage{textcomp,amsmath,graphicx,amssymb}
\usepackage{multirow,array} 
\usepackage{float}
\usepackage{yfonts,mathrsfs} 
\usepackage{color, colortbl}
\usepackage{multirow}
\usepackage{tabularx}
\usepackage{makecell}
\usepackage{graphicx}
\usepackage[caption=false]{subfig}
\usepackage{algorithm}
\usepackage{algorithmic}
\usepackage{authblk}
\usepackage{latexsym}

\graphicspath{{./figures/}}

\firstpage{1} 
\pubvolume{xx}
\issuenum{1}
\articlenumber{5}
\pubyear{2018}
\copyrightyear{2018}
\history{First version: September, 2018}

\pdfoutput=1

\Title{Pricing options and computing implied volatilities using neural networks}

\Author{Shuaiqiang Liu $^{1,}$*, Cornelis W. Oosterlee $^{1,2}$ and Sander M.Bohte $^{2}$}

\AuthorNames{Shuaiqiang Liu, Cornelis W. Oosterlee and Sander M.Bohte}

\address{%
$^{1}$ \quad Applied Mathematics (DIAM), Delft University of Technology, Building 28,
Mourik Broekmanweg 6, 2628 XE, Delft, Netherlands; \\
$^{2}$ \quad Centrum Wiskunde \& Informatica, Science Park 123, 1098 XG, Amsterdam, Netherlands;}

\corres{Correspondence: s.liu-4@tudelft.nl}

\abstract{This paper proposes a data-driven approach, by means of an Artificial Neural Network (ANN), to value financial options and to calculate implied volatilities with the aim of accelerating the corresponding numerical methods. With ANNs being universal function approximators, this method trains an  optimized ANN on a data set generated by a sophisticated financial model, and runs the trained ANN as an agent of the original solver in a fast and efficient way. We test this approach on three different types of solvers, including the analytic solution for the Black-Scholes equation, the COS method for the Heston stochastic volatility model and Brent's iterative root-finding method for the calculation of implied volatilities. The numerical results show that the ANN solver can reduce the computing time significantly.}


\begin{document}

\section{Introduction}

In computational finance, numerical methods are commonly used for the valuation of financial derivatives and also in modern risk management. Generally speaking, advanced financial asset models are able to capture nonlinear features that are observed in the financial markets. However, these asset price models are often multi-dimensional, and, as a consequence, do not give rise to closed-form solutions for option values. 

Different numerical methods have therefore been developed to solve the corresponding option pricing partial differential equation (PDE) problems, e.g. finite differences, Fourier methods and Monte Carlo simulation. 
In the context of financial derivative pricing, there is a stage in which the asset model needs to be calibrated to market data. In other words, the open parameters in the asset price model need to be fitted. This is typically {\em not} done by historical asset prices, but by means of {\em option prices}, i.e. by matching the market prices of heavily traded options to the option prices from the mathematical model, under the so-called risk-neutral probability measure. In the case of model calibration, thousands of option prices need to be determined in order to fit these asset parameters. 
However, due to the requirement of a highly efficient computation,  certain high quality asset models are discarded. 
Efficient numerical computation is also increasingly important in financial risk management, especially when we deal with real-time risk management (e.g., high frequency trading) or counterparty credit risk issues, where a trade-off between efficiency and accuracy seems often inevitable.

Artificial neural networks (ANNs) with multiple hidden layers have become successful machine learning methods to extract features and detect patterns from a large data set. There are different neural network variants for particular tasks, for example, convolutional neural networks for image  recognition and recurrent neural networks for time series analysis. It is well-known that ANNs can approximate nonlinear functions \cite{Cybenko1989}, \cite{Hornik1991ApproximationNetworks}, \cite{Hornik1990UniversalNetworks}, and can thus be used to approximate solutions to PDEs \cite{Lagaris1998ArtificialEquations}, \cite{2018DGM_NN}. Recent advances in data science have shown that using deep learning techniques even highly nonlinear multi-dimensional functions can be accurately represented~\cite{LeCun2015DeepLearning}. Essentially, ANNs can be used as powerful universal function approximators without assuming any mathematical form for the functional relationship between the input variables and the output. Moreover, ANNs easily allow for parallel processing to speed up evaluations, especially on GPUs.

We aim to take advantage of a classical ANN to speed up option valuation by learning the results of an option pricing method. From a computational point of view, the ANN does not suffer much from the dimensionality of a PDE.  An ``ANN solver'' is typically decomposed into two separate phases, a training phase and a test (or prediction) phase. During the training phase, the ANN ``learns'' the PDE solver, by means of the data set generated by the sophisticated models and corresponding numerical solvers. This stage is usually time consuming, however, it can be done off-line. During the test phase, the trained model can be employed  to approximate the solution on-line.  The ANN solution can typically be computed as a set of matrix multiplications, which can be implemented in parallel and highly efficiently, especially with GPUs.  As a result, the trained ANN delivers financial derivative prices, or other quantities, efficiently, and the on-line time for accurate option pricing may be reduced, especially for involved asset price models. We will show in this paper that this data-driven approach is highly promising.

The proposed approach in this paper attempts to accelerate the pricing of European options under a unified data-driven ANN framework.  ANNs have been used in option pricing for some decades already. There are basically two directions. One is that based on the observed market option prices and the underlying asset value, ANN-based regression techniques have been applied to fit a model-free, non-parametric pricing function, see, for example, \cite{HUTCHINSON1994ANetworksb}, \cite{Yao2000OptionNetworks}, \cite{2001PricingandHedgingNN}, \cite{Garcia2000PricingHint}. Furthermore, the authors of \cite{2000SecondOrderNNOptionPricing}, \cite{Yang2016GatedDesign} designed special kernel functions to incorporate prior financial knowledge into the neural network while forecasting option prices.  

Another direction is to improve the  performance of model-based pricing by means of ANNs. The interest in accelerating classical PDE solvers via ANNs is rapidly growing. The papers \cite{Han2017OvercomingLearning}, \cite{E2017DeepEquations}, \cite{Beck2017MachineEquations} take advantage of reinforcement learning to speed up solving high-dimensional stochastic differential equations. The author of \cite{2017SGD4AmericaOption} proposes  an optimization algorithm, the so-called  stochastic gradient descent in continuous time, combined with a deep neural network to price high-dimensional American options. In \cite{Fan2009OptionMethods} the pricing performance of financial models is enhanced by  non-parametric learning approaches that deal with a systematic bias of pricing errors. Of course, this trend takes place not only in computational finance, but also in other engineering fields where PDEs play a key role, like computational fluid dynamics, see \cite{HESTHAVEN201855},  \cite{Raissi2018DeepEquations}, \cite{2018DGM_NN}, \cite{Tompson2016AcceleratingNetworks}.
The work in this paper belongs to this latter direction. Here, we use traditional solvers to generate artificial data, then we train the ANN to learn the solution for different problem parameters.  Compared to \cite{Lagaris1998ArtificialEquations} or \cite{2018DGM_NN}, our data-driven approach finds, next to the solutions of the option pricing PDEs, the implicit relation between variables and a specific parameter (i.e. the implied volatility).

This paper is organized as follows.  In section \ref{section:Option pricing}, two fundamental option pricing models, the Black-Scholes and the Heston stochastic volatility PDEs, are briefly introduced. In addition to European option pricing, we also analyze robustness issues of root-finding methods to compute the so-called implied volatility. In section \ref{section:Methodology}, the employed ANN is presented with suitable hyper-parameters. After training the ANN to learn the results of the financial models for different problem parameters, numerical ANN results with the corresponding errors are presented in section \ref{section:Numerical Results}.


\section{Option pricing and asset models} \label{section:Option pricing}

In this section, two asset models are briefly presented, the geometric Brownian motion (GBM) asset model, which gives rise to the Black-Scholes option pricing PDE, and the Heston stochastic volatility asset model, leading to the Heston PDE. We also discuss the concept of implied volatility.  We will use European option contracts as the examples, however, other types of options can be taken into consideration in a similar way. 

\subsection{The Black-Scholes PDE}

A first model for asset prices is GBM,

\begin{equation}\label{eq:gbm}
dS_t = \mu S_t dt + \sqrt{\nu} S_t dW^s_t,
\end{equation}
where $S$ is the price of an non-dividend paying asset, and $W^s$ is a Wiener process, with $t$ being the time,  $\mu$ the drift parameter, and $\nu$  the variance parameter. The volatility parameter is $\sigma=\sqrt{\nu}$. A European option contract on the underlying stock price can be valued via the Black-Scholes PDE, which can be derived from  It\^{o}'s Lemma under a replicating portfolio approach or via the martingale approach. Denoting the option price by $V(t,S)$, the Black-Scholes equation reads,

\begin{equation} \label{eq:bs-pde}
\frac{\partial V}{\partial t} + \frac{1}{2}\sigma^2 S^2 \frac{\partial^2 V}{\partial S^2} + rS\frac{\partial V}{\partial S} - rV = 0,
\end{equation}
with time $t$ until to maturity $T$, and $r$ the risk-free interest rate. The PDE is accompanied by a final condition representing the specific payoff, for example, the European call option payoff at time $T$,

\begin{equation}
V(t=T,S)= (S_0-K)^+,
\label{po}
\end{equation}
where $K$ is the option's strike price. See standard textbooks for more information about the basics in financial mathematics.

An analytic solution to~(\ref{eq:bs-pde}),~(\ref{po}) exists for European plain vanilla options, i.e., 
\begin{subequations}  \label{eq:bs_solution}
\begin{align} 
V_c(t,S) &= S N(d_1) -Ke^{-r\tau}N(d_2),\\
d_1 &= \frac{\log(S/K)+(r-0.5\sigma^2)\tau}{\sigma \sqrt{\tau}}, \;\;  d_2 = d_1- \sigma \sqrt{\tau},
\end{align}
\end{subequations}
where $\tau:=T-t$, $V_c(t,S)$ is the European call option value at time $t$ for stock value $S$, and $N(\cdot)$ represents the normal distribution. This solution procedure~(\ref{eq:bs_solution}) is denoted by $V(\cdot)=BS(\cdot)$.


\subsubsection{Implied volatility} \label{section:iv}

Implied volatility is considered an important quantity in finance. Given an observed market option price $V^{mkt}$, the Black-Scholes implied volatility $\sigma^*$ can be determined by solving $BS(\sigma^*; S,K,\tau,r)=V^{mkt}$. The monotonicity of the Black-Scholes equation with respect to the volatility guarantees the existence of $\sigma^* \in [0,+\infty]$. We can write the implied volatility as an implicit formula, 

\begin{equation} \label{eq:ivf}
\sigma^*(K,T) = BS^{-1}(V^{mkt};S,K,\tau,r),
\end{equation}
where $BS^{-1}$ denotes the inverse Black-Scholes function. Moreover, by adopting moneyness, $m=\frac{S_t}{K}$, and time to maturity, $\tau=T-t$, one can express  the implied volatility as $\sigma^*(m,\tau)$, see  \cite{Cont2002DynamicsSurfacesb}.

 For simplicity, we denote  here $\sigma^*(m,\tau)$ by  $\sigma^*$.  An analytic solution for Equation (\ref{eq:ivf}) does not exist. The value of $\sigma^*$ is determined by means of a numerical iterative technique, since  Equation (\ref{eq:ivf})  can be converted into a root-finding problem, 
 
\begin{equation} \label{eq:iv}
 g(\sigma^*)=BS(S,\tau,K,r,\sigma^*)-V^{mkt}(S,\tau;K)=0. 
\end{equation}


\subsection{The Heston Model}

One of the limitations of using the Black-Scholes model is the assumption of a constant volatility $\sigma$ in~(\ref{eq:bs-pde}),~(\ref{eq:bs_solution}). A major modeling step away from the assumption of constant volatility in asset pricing, was made by 
modeling the {\it volatility/variance as a diffusion process}. 
The resulting models are the stochastic volatility (SV) models. 
The idea to model volatility as a random variable is confirmed by practical financial data which indicates the variable and unpredictable nature of the stock price's volatility. 
The most significant argument to consider the volatility to be stochastic is the implied volatility smile/skew, which is present in the financial market data, and can be accurately recovered by SV models, especially for options with a medium to long time to the maturity date $T$. With an additional stochastic process, which is correlated to the asset price process $S_t$,
we deal with a {\em system of SDEs}, for which option valuation is more expensive than for a scalar asset price process.

The most popular SV model is the Heston model \cite{Heston1993AOptions}, for which the system of stochastic equations under the risk-neural measure reads, 

\begin{subequations}  \label{eq:heston}
\begin{align}
& dS_t   = r S_t dt + \sqrt{\nu_t} S_t dW^s_t, S_{t_0}=S_0,\\
& d\nu_t = \kappa(\bar{\nu} - \nu_t) dt + \gamma \sqrt{\nu_t}dW^{\nu}_t, \nu_{t_0}=\nu_0, \\
& dW^s_t  dW^{\nu}_t = \rho dt,
\end{align}
\end{subequations}
with $\nu_t$ the instantaneous variance, and $W^s_t, W^{\nu}_t$ are two Wiener processes with correlation coefficient $\rho$. The second equation in~(\ref{eq:heston}) models a mean reversion process for the variance, with the parameters, $r$ the risk-free interest rate, $\bar{\nu}$ the long term variance, $\kappa$ the reversion speed; $\gamma$ is the volatility of the variance, determining the volatility of $\nu_t$. There is an additional parameter $\nu_0$, the $t_0$-value of the variance.

By the martingale approach, we arrive at the following multi-dimensional Heston option pricing PDE,
\begin{eqnarray}
        \frac{\partial V}{\partial t}&+&rS\frac{\partial
V}{\partial S}+\kappa(\bar{\nu}-\nu)\frac{\partial V}{\partial \nu} 
+\frac12 \nu S^2\frac{\partial^2 V}{\partial S^2} \nonumber \\
        &+& \rho\gamma S
\nu \frac{\partial^2V}{\partial S\partial \nu}+\frac12\gamma^2
\nu\frac{\partial^2 V}{\partial \nu^2}-r{V}=0.
\label{ch1:hmpo}
\end{eqnarray}
The typically observed implied volatility shapes in the market, e.g. smile or skew, can be reproduced by varying the above parameters $\{\kappa, \rho, \gamma, \nu_0, \bar{\nu}\}$. In general, the parameter $\gamma$ impacts the kurtosis of the asset return distribution, and the coefficient $\rho$ controls its asymmetry. The Heston model does not have analytic solutions, and is thus solved numerically.

Numerical methods in option pricing generally fall into three categories, finite differences (FD), Monte Carlo (MC) simulation and numerical integration methods.  Finite differences for the PDE problem are often used for free boundary problems, as they occur when valuing American options, or for certain exotic options like barrier options. The derivatives of the option prices (the so-called option Greeks) are accurately computed with finite differences.

Monte Carlo simulation and numerical integration rely on the Feyman-Kac Theorem, which essentially states that (European) option values can be written as discounted expected values of the option's payoff function at the terminal time $T$, under the risk-neutral measure. Monte Carlo methods are often employed in this context for the valuation of path-dependent high-dimensional options, and also for the computation of all sorts of valuation adjustments in modern risk management.  However, Monte Carlo methods are typically somewhat slow to converge, and particularly in the context of model calibration this can be an issue.

The numerical integration methods are also based on the Feyman-Kac Theorem. The preferred way to employ them is to first transform to the Fourier domain. 
The availability of the asset price's characteristic function is a pre-requisite to using Fourier techniques. 
One of the efficient techniques in this context is the COS method \cite{Fang20091.Expansions}, which  utilizes Fourier-cosine series expansions to approximate the asset price's probability density function, but is based on the characteristic function. 
The COS method can be used to compute European option values under the Heston model highly efficiently. 
However, for many different, modern asset models the characteristic function is typically not available. 
We will use the Heston model with the COS method here during the training of the Heston-ANN, so that training time is still relatively small.

\subsection{Numerical methods for implied volatility}
Focussing on the implied volatility $\sigma^*$, there are several iterative numerical techniques to solve~(\ref{eq:iv}), like, for example, the Newton-Raphson method, the bisection method or the Brent method. 
The Newton-Raphson iteration reads,

\begin{equation}
\sigma^*_{k+1}=\sigma^*_{k}- \frac{V(\sigma^*_k)-V^{mkt}}{ g'(\sigma^*_{k})}, \; k=0,\ldots. 
\label{eq:newton}
\end{equation}

Starting with an initial guess, $\sigma^*_{0}$, the approximate solutions, $\sigma^*_{k+1}$, $k=0,\ldots$, iteratively improve, until a certain criterion is satisfied. 
The first derivative of Black-Scholes option value with respect to the volatility, named the option's Vega, in the denominator of~(\ref{eq:newton}) can be obtained analytically for European options. 

\begin{figure}[H]
    \centering
    \subfloat[Option price vs. volatility] {\label{fig:option_vol_curve_big}    
        \includegraphics[width=10cm]{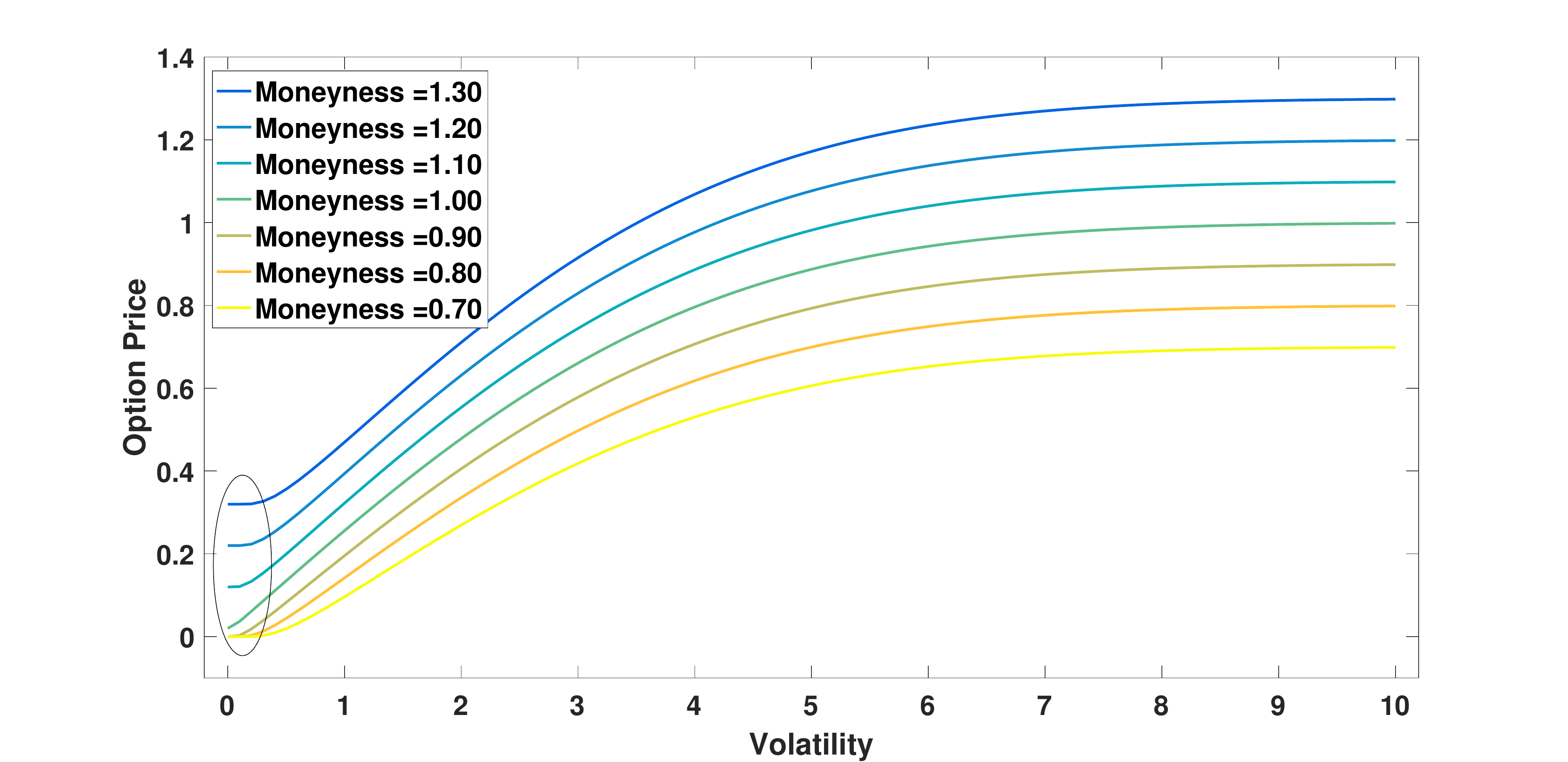} }  
\hfill
      \subfloat[Vega vs. Moneyness] {\label{fig:vega_call}   
        \includegraphics[width=10cm]{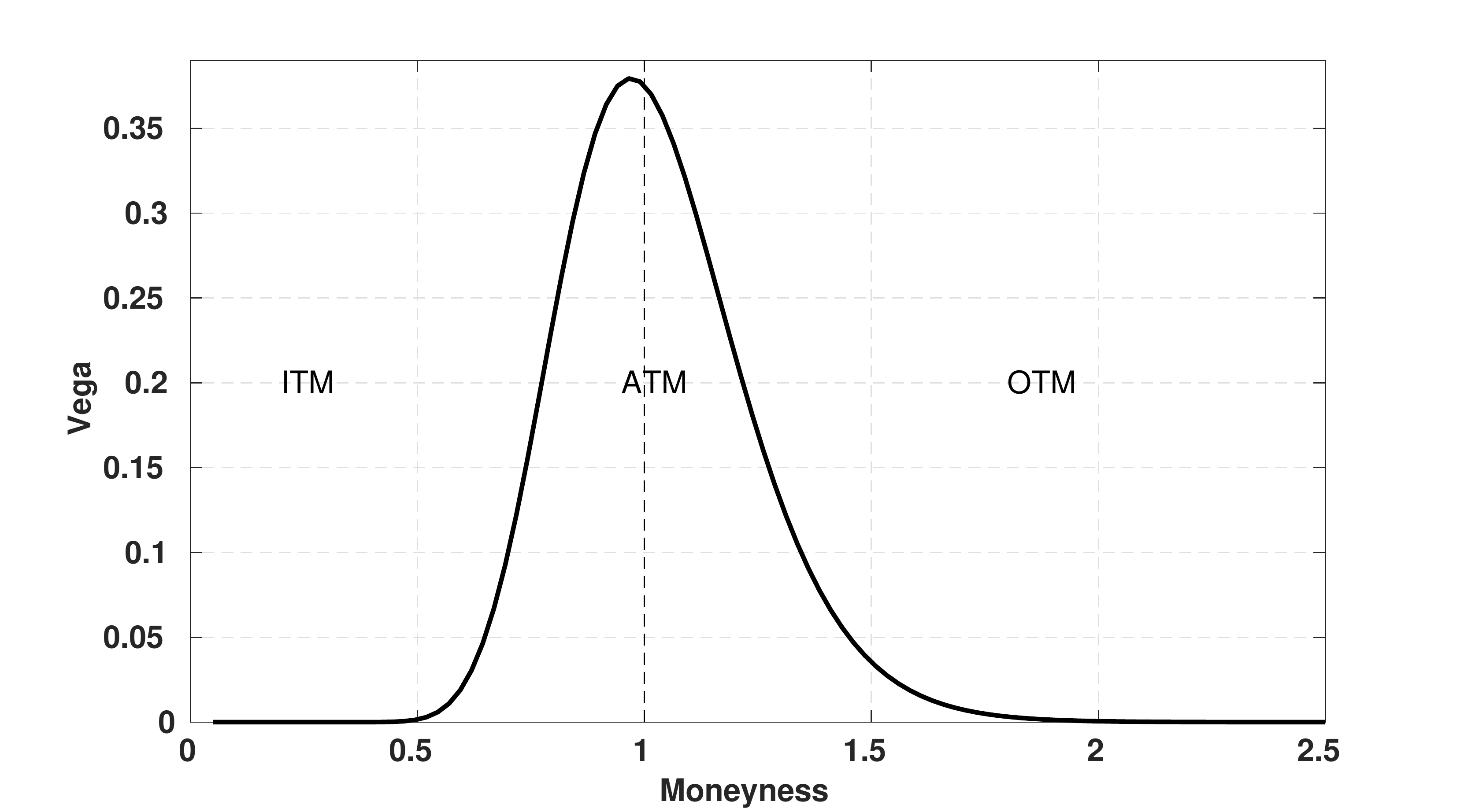} }
    \caption{Vega tends to be zero in certain regions of deep ITM or OTM options.}    
    \label{fig:option_vol_curve}
\end{figure}

However, the Newton-Raphson method may fail to converge, either when the Vega is extremely small or when convergence stalls.  The Black-Scholes equation monotonically maps an unbounded interval $\sigma \in [0,+\infty)$ to a finite range $V(t,S) \in [0,S_t-Ke^{-r\tau}]$, and the option Vega can be very close to zero  in certain $\sigma$-regions, especially when the option is either deep in-the-money (ITM) or deep out-the-money (OTM). Figure \ref{fig:vega_call} shows that Vega is relatively large in the at-the money (ATM) region, but the near-flat function shapes appear in the regions with small or large volatilities of deep ITM or OTM options. A possible robust root-finding alternative to this problem is to employ a hybrid of the Newton-Raphson and the bisection methods. Alternatively, the author of \cite{Jackel2015LetsRational} proposed to select a suitable initial value at the beginning of the iteration to avoid divergence. In the next subsection, we will discuss a derivative-free, robust and efficient algorithm to find the implied volatility. 

\subsubsection{Brent's method for implied volatility}

As a derivative-free, robust and efficient algorithm,  Brent's method \cite{brentmethod} combines bisection, inverse quadratic interpolation and the secant method. In order to determine the next iterant, an inverse quadratic interpolation employs three prior points (i.e. iterants) to fit an inverse quadratic function, which resembles the gradient of Newton's method, i.e.

\begin{equation}
	\begin{aligned}
\sigma_{k+1} & =  \frac{\sigma_{k} g(\sigma_{k-1}) g(\sigma_{k-2})}{(g(\sigma_{(k)})-g(\sigma_{k-1}))(g(\sigma_{k})-g(\sigma_{k-2}))} \\
& + \frac{\sigma_{k-1} g(\sigma_{k-2}) g(\sigma_{k})}{(g(\sigma_{k-1})-g(\sigma_{k-2}))(g(\sigma_{k-1})-g(\sigma_{k}))} \\
& + \frac{\sigma_{k-2} g(\sigma_{k-1}) g(\sigma_{k})}{(g(\sigma_{k-2})-g(\sigma_{k-1}))(g(\sigma_{k-2})-g(\sigma_{k}))}.
\end{aligned}
\end{equation}

When two consecutive approximations are identical, for example, $\sigma_{k}=\sigma_{k-1}$, the quadratic interpolation is replaced by an approximation based on the secant method,

\begin{equation}
\sigma_{k+1}=\sigma_{k-1} -g(\sigma_{k-1}) \frac{\sigma_{k-1}-\sigma_{k-2}}{g(\sigma_{k-1})-g(\sigma_{k-2})}.
\end{equation}

In this paper, Brent's method is used to compute the BS implied volatility related to the Heston option prices in Section \ref{section:heston-iv}.
We will develop an ANN to approximate the implicit function relating the volatility to the option price.


\section{Methodology} \label{section:Methodology}

In this section, we present a neural network to approximate a function for financial models.  The procedure comprises two main components, the generator to create the financial data for training the model and the predictor (the ANN) to approximate the option prices based on the trained model.  The data-driven framework consists of the following steps,

\begin{algorithm}[H]
\setlength\baselineskip{18pt}
\caption{Model framework }
\begin{algorithmic} 
\STATE -- Generate the sample data points for input parameters, 
\STATE -- Calculate the corresponding output (option price or implied volatility) to form a complete data set with inputs and outputs,
\STATE -- Split the above data set into a training and a test part,
\STATE -- Train the ANN on the training data set, 
\STATE -- Evaluate the ANN on the test data set,
\STATE -- Replace the original solver by the trained ANN in applications.
\end{algorithmic}
\end{algorithm}


\subsection{Artificial neural network} \label{sec:dnn}

ANNs generally constitute three levels of components, i.e. neurons, layers and the architecture from bottom to top. The architecture is determined by a combination of different layers, that are made up of numerous artificial neurons. A neuron, which involves learnable weights and biases, is the fundamental unit of ANNs. By connecting the neurons of adjacent layers, output signals of a previous layer enter a next layer as input signal. By stacking layers on top of each other, signals travel from the input layer through the hidden layers to the output layer potentially through cyclic or recurrent connections, and the ANN builds a mapping among input-output pairs. 

\begin{figure}[H]
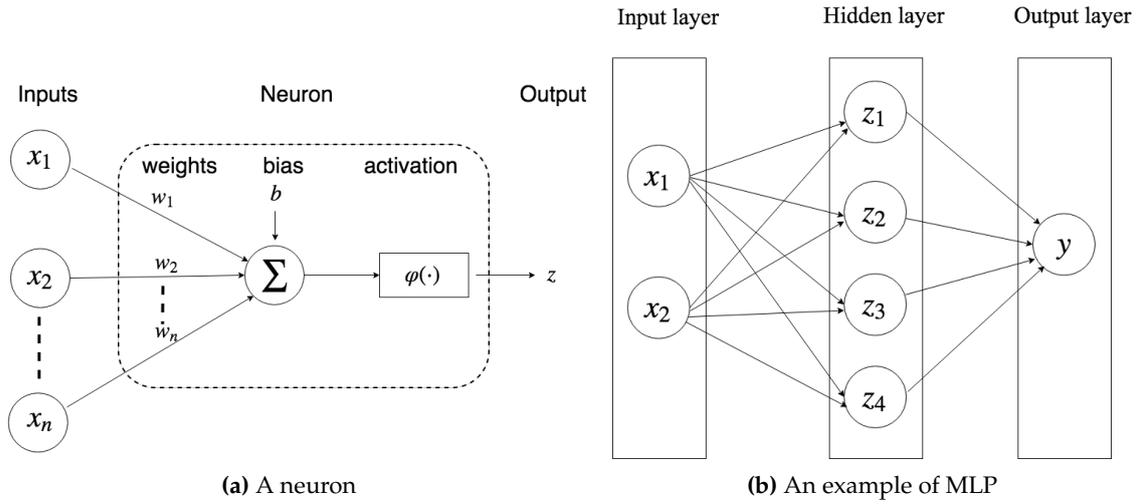

    \centering
    \subfloat[A neuron] {\label{fig:neuron}
        \includegraphics[width=8cm]{neuron} }  
      \subfloat[An example of MLP] {\label{fig:mlp_structure}     
        \includegraphics[width=7cm]{mlp_structure} }
    \caption{Illustration of an MLP configuration.}    
    \label{fig:neuron_structure}
\end{figure}

As shown in Figure \ref{fig:neuron}, an artificial neuron  basically consists of the following three consecutive operations:
\begin{enumerate}[leftmargin=*,labelsep=4.9mm]
\item Calculation of a summation of weighted inputs,
\item Addition of a bias to the summation,
\item Computation of the output by means of a transfer function.
\end{enumerate}

A multi-layer perceptron (MLP), consisting of at least three layers is the simplest version of an ANN.
Mathematically, MLPs are defined by the following parameters,

 \begin{equation}
 \boldsymbol{\theta} = (\mathbf{W}_1, \mathbf{b}_1, \mathbf{W}_2, \mathbf{b}_2,..., \mathbf{W}_L,\mathbf{b}_L)
 \end{equation}
where $\mathbf{W}_j$ is a weight matrix and  $\mathbf{b}_j$ is the bias vector of the $L$-th neural layer. A function can then be expressed as follows,

\begin{equation} \label{eq:fun-dnn}
y( \mathbf{x}) = F( \mathbf{x} |\boldsymbol{\theta}).
\end{equation}

Let $z_{j}^{(l)}$ denote the value of the $j$-th neuron in the $l$-th layer, then the corresponding transfer function reads,

 \begin{equation}\label{eq:zz}
 z_{j}^{(l)}=\varphi^{(l)} \left(\sum_{i}w_{ij}^{(l)}z^{(l-1)}_{i}+b_{j}^{(l)} \right),
\end{equation}
where $z_i^{(l-1)}$ is the output value of the $i$-th neuron in the ($l$-1)-th layer  and $\varphi(\cdot)$ is an activation function, with $w_{ij}^{(l)} \in \mathbf{W}_l$, $b_j^{(l)} \in \mathbf{b}_l$. When $l$=0, $z^{(0)}=x$ is the input layer; When $l$=$L$, $ z^{(L)}=y$ is the output layer; Otherwise, $z^{(l)}$  represents an intermediate variable. The activation function $\varphi(\cdot)$ adds non-linearity to the system, for example, the following activation functions may be employed,
\begin{itemize}[leftmargin=*,labelsep=5.8mm]
\item Relu, $\varphi(x) = \max(x,0)$,
\item Sigmoid, $\varphi(x) = \frac{\displaystyle 1}{\displaystyle 1+e^{-x}}$,
\item Leaky ReLu, $\varphi(x) = \max(x,ax)$, $0<a<1$;
\end{itemize}
see \cite{LeCun2015DeepLearning} for more activation functions. Equation (\ref{eq:nn-formula-one}) presents an example of the formula of an MLP with ``one-hidden-layer'',

\begin{equation}\label{eq:nn-formula-one} 
\begin{cases} 
y = \varphi^{(2)} \left(\sum_{j}w_{j}^{(2)}z^{(1)}_{j}+b^{(2)} \right) \\
z_{j}^{(1)} = \varphi^{(1)} \left(\sum_{i}w_{ij}^{(1)}x_{i}+b_{j}^{(1)} \right).
\end{cases}
\end{equation}

According to the Universal Approximation Theorem \cite{Cybenko1989}, a single-hidden-layer ANN with a sufficient number of neurons can approximate any continuous function. The distance between two functions is measured by the norm of a function $||\cdot||$,

\begin{equation}
D(f(\mathbf{x}),F(\mathbf{x})) = || f(\mathbf{x})- F(\mathbf{x}) ||,
\end{equation}
where $f(\mathbf{x})$ is the objective function, $F(\mathbf{x})$ is the neural network approximated function.  For example, the $p$-norm reads,

$$||f(\mathbf{x}) - F(\mathbf{x}|\boldsymbol{\theta})||_p=\sqrt[p]{\int_\mathbf{X}|f(\mathbf{x})-F(\mathbf{x}|\boldsymbol{\theta})|^p d \mu(\mathbf{x})}, $$ 
where $1 \leq p < \infty$  and  $\mu(\mathbf{x})$ is a measure over the space $\mathbf{X}$.  We choose $p$=2 to evaluate the averaged accuracy, which corresponds to the mean squared error (MSE). Within supervised learning,  the loss function is equivalent to the above distance,

\begin{equation}
L(\boldsymbol{\theta}):=D(f(\mathbf{x}), F(\mathbf{x}|\boldsymbol{\theta})).
\end{equation}

The training process aims to learn the optimal weights and biases in Equation (\ref{eq:fun-dnn}) to make the loss function as small as possible. The process can be formulated as an optimization problem,

\begin{equation} \label{eq:argmin_dnn}
arg\min_{\boldsymbol{\theta}} L(\boldsymbol{\theta} | (\mathbf{x},\mathbf{y})),
\end{equation}
given the known input-output pairs $(\mathbf{x},\mathbf{y})$ and a loss function $L(\boldsymbol{\theta})$.  

A number of back-propagation gradient descent methods \cite{review_sgd2016} have been successfully applied to solve Equation (\ref{eq:argmin_dnn}), for instance, Stochastic Gradient Descent (SGD) and its variants Adam and RMSprop. These optimization algorithms start with initial values and move in the direction in which the loss function decreases. The formula for updating the parameters reads,

\begin{equation}\label{eq:nn-formula}
\begin{cases} 
\mathbf{W} \leftarrow  \mathbf{W} - \eta(i) \frac{\displaystyle \partial L}{\displaystyle  \partial \mathbf{W}}, \\[1.5ex] 
\mathbf{b} \leftarrow  \mathbf{b} - \eta(i) \frac{\displaystyle \partial L}{\displaystyle  \partial \mathbf{b}},\\[1.5ex]
i=0,1,2, ... ,\\ 
\end{cases}
\end{equation}
where $\eta$ is a learning rate, which may vary during the iterations. The learning rate plays an important role during the training, as a ``large'' learning rate value causes the ANN's convergence to oscillate, whereas a small one results in ANNs learning slowly, and even get trapped in local optima regions. An adaptive learning rate is often preferred, and more detail will be given in Section \ref{section:learning_rate}.

\subsection{Hyper-parameters optimization}

Training deep neural networks involves numerous choices for the commonly called  ``hyper-parameters''. These include the number of layers, neurons, and the specific activation function.  Determining the depth (the number of hidden layers) and the width (the number of neurons) of the ANN is a challenging problem.

We experimentally find that  an MLP architecture with four hidden layers has an optimal capacity of approximating option pricing formulas of our current interest. Built on a four hidden layer architecture, the other hyper-parameters are optimized using automatic machine learning~\cite{BergstraAlgorithmsOptimization}. There are different techniques to implement the automatic search. In a grid search technique, all candidate parameters are systematically parameterized on a pre-defined grid, and all possible candidates are explored in a brute-force way. The authors of~\cite{Bergstra2012RandomBengio} concluded that random search is more efficient for hyper-parameters optimization. Recently, Bayesian hyper-parameter optimization has been developed to efficiently reduce the computational cost by navigating through the hyper-parameters space. 
However, it is difficult to outperform random search in combination with certain expert knowledge.

Neural networks may not necessarily converge to a global minimum. However, using a proper {\em random initialization} may help the model with suitable initial values. {\em Batch normalization} scales the output of a layer by subtracting the batch mean and dividing by the batch standard deviation. This  can speed up the training of the neural network.
The batch size indicates the number of samples that enter the model to update the learnable parameters within one iteration. 
A {\em dropout operation} selects a random set of neurons and deactivates them, which forces the network to learn more robust features. The dropout rate refers to the proportion of the deactivated neurons in a layer. 

There are two stages to complete the hyper-parameter optimization. During the model selection process, over-fitting can be reduced by adopting the $k$-fold cross validation as follows.  
\begin{algorithm}[H]
\setlength\baselineskip{18pt}
\caption{ $k$-fold cross validation }
\begin{algorithmic}[leftmargin=*,labelsep=5.8mm] 
\STATE --Split the training data set into $k$ different subsets,
\STATE --Select one set as the validation data set,
\STATE --Train the model on the remaining $k$-1 subsets,
\STATE  --Calculate the metric by evaluating the trained model on the validation part,
\STATE  --Continue the above steps by exploring all subsets,
\STATE  --Calculate the final metric which is averaged over $k$ cases,
\STATE  --Explore the next set of hyper-parameters,
\STATE  --Rank the candidates according to their averaged metric.
\end{algorithmic}
\end{algorithm}

\begin{table}[htp]
\begin{center}
\caption{The setting of random search for hyper-parameters optimization} 
  \begin{tabular}{  c | c  }
    \hline

    Parameters          & Options or Range \\ \hline

    Activation          & ReLu, tanh, sigmoid, elu \\

    Dropout rate 		& [0.0, 0.2] \\ 

    Neurons             & [200, 600] \\

    Initialization      & uniform, glorot\_uniform, he\_uniform\\

    Batch normalization & yes, no \\

    Optimizer           & SGD, RMSprop, Adam \\

    Batch size          & [256, 3000]  \\

    \hline	  

  \end{tabular}
  \label{table:Random search}
 \end{center}
  \end{table}
In the first stage, we employ random search combined with a 3-fold cross validation to find initial hyper-parameter configurations for the neural network. As shown in Table \ref{table:Random search}, each model is trained 200 epochs using MSE as the loss metric.  
An epoch is the moment when the model has processed the whole training data set.  
The prediction accuracy increases  with  the training data set size (more details will be discussed in Section \ref{section: data set impact}). The random search is implemented on a small data set, which is then followed by training the selected ANN on larger data sets in the next stage.

 \begin{table}[htp]

\begin{center}
\caption{The selected model after the random search}

  \begin{tabular}{  c | c  }

    \hline

    Parameters          &  Options  \\ \hline    

    Hidden layers & 4  \\

    \hline

    Neurons(each layer) &  400 \\

    Activation          & ReLu \\

    Dropout rate 		& 0.0 \\ 

    Batch-normalization  & No \\

    Initialization      &  Glorot\_uniform\\

    Optimizer           &  Adam \\

    Batch size          & 1024  \\

    \hline	  

  \end{tabular}

   \label{table:final-nn-setting}

 \end{center}

  \end{table}

 In the second stage,  we further enhance the top 5 network configurations by averaging the different values, to yield the final ANN model, as listed in Table \ref{table:final-nn-setting}. As Table \ref{table:final-nn-setting} shows, the optimal parameter values do not lie at the boundaries of the search space (except for the drop out rate). Batch normalization and drop-out do not improve the model accuracy in this regression problem, and one possible reason is that the output value is sensitive to the input parameters, which is different from  sparse features in an image (where these operations usually work very well). Subsequently, we train the selected network on the whole (training and validation) data set, to obtain the final weights.  This procedure results in an ANN with sufficient accuracy to approximate the financial option values.

\subsection{Learning rates} \label{section:learning_rate}

The learning rate, one of the key hyper-parameters, represents the rate at which the weights are updated each iteration. A large learning rate leads to fluctuations around a local minimum, and sometimes even to divergence. Small learning rates may cause an inefficiently slow training stage. It is common practice to start with a large learning rate and then gradually decrease it until a well-trained model has resulted.  There are different ways to vary the learning rate during training, e.g.  by step-wise annealing, exponential decay, cosine annealing, see \cite{SmithCyclicalNetworks} for a cyclical learning rate (CLR) and \cite{Loshchilov2016SGDR:Restarts} for the stochastic descent gradient restart (SDGR). The  basic idea of CLR and SDGR is that at certain points of the training stage, a relatively large learning rate may move the weights from their current values, by which ANNs may leave a local optimum and converge to a better one.

\begin{figure}[H]
\centering
\includegraphics[width = 0.6\textwidth]{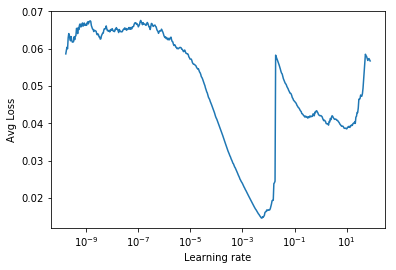}
\caption{ Average training loss against varying learning rates.}
\label{fig:loss_vs_learning_rate}
\end{figure}

We employ the method proposed in \cite{SmithCyclicalNetworks} to determine the learning rate. The method is based on the insight of how the averaged training loss varies over different learning rates, by starting with a small learning rate and increasing it progressively in the first few iterations. By monitoring the loss function against the learning rate, it is shown in Figure \ref{fig:loss_vs_learning_rate} that the loss stabilizes when the learning rate is small, then drops rapidly and finally oscillates and diverges when the learning rate is too large. The optimal learning rate lies here between $10^{-5}$ and $10^{-3}$, where the slope is the steepest and the training loss reduces quickly. Therefore, the learning rate in CLR is reduced from $10^{-3}$  to $10^{-5}$ in our experiments. 

\begin{figure}[H]
\centering
\includegraphics[width = 0.7\textwidth]{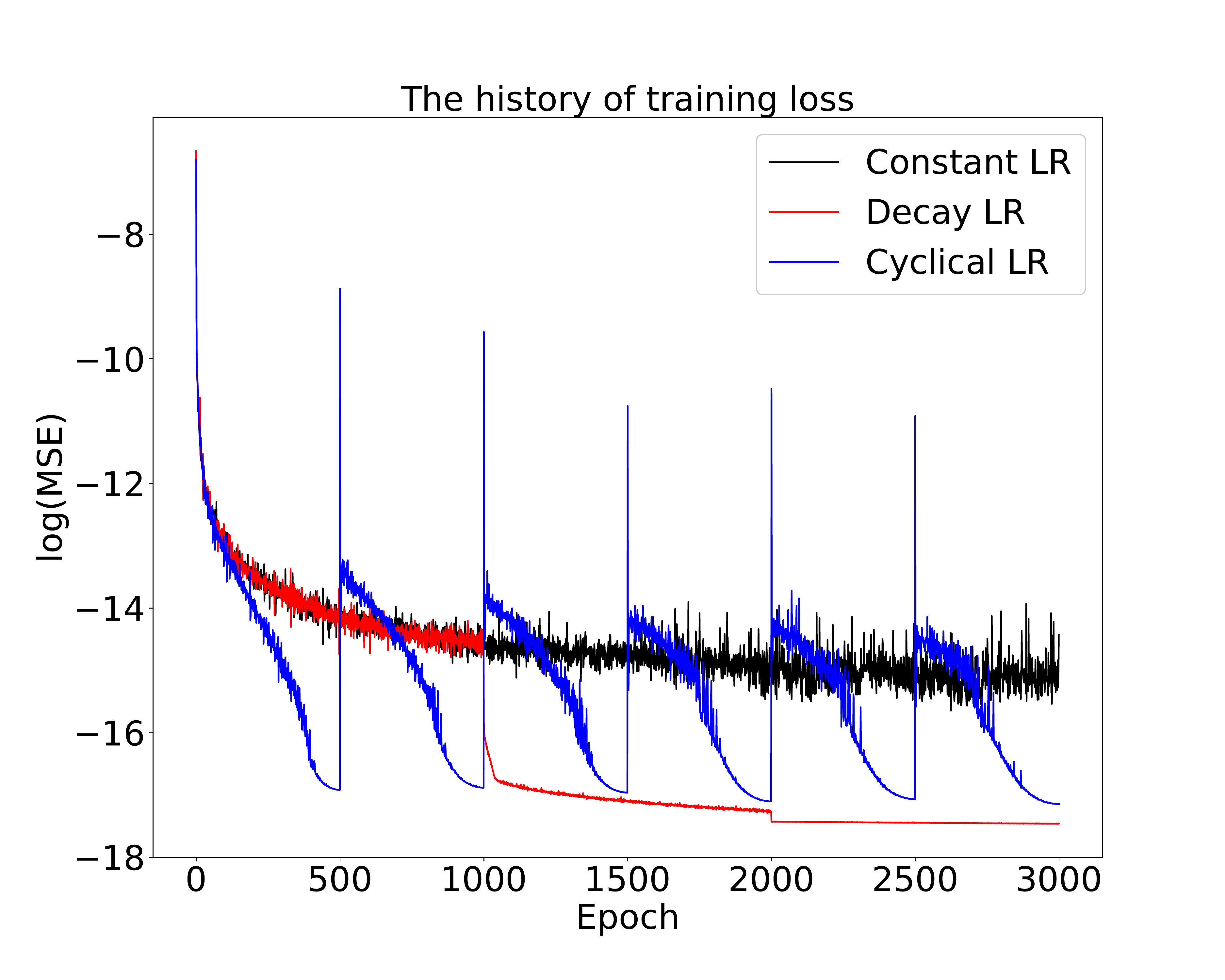}
\caption{ Different learning rate schedules for training  ANNs  on  Heston model.}
\label{fig:nn_training_loss_comparison}
\end{figure}

\begin{figure}[H]
    \centering
    \subfloat[Losses with decaying learning rates] {\label{fig:decay_LR_train_valuation_loss_heston}
        \includegraphics[width=8cm]{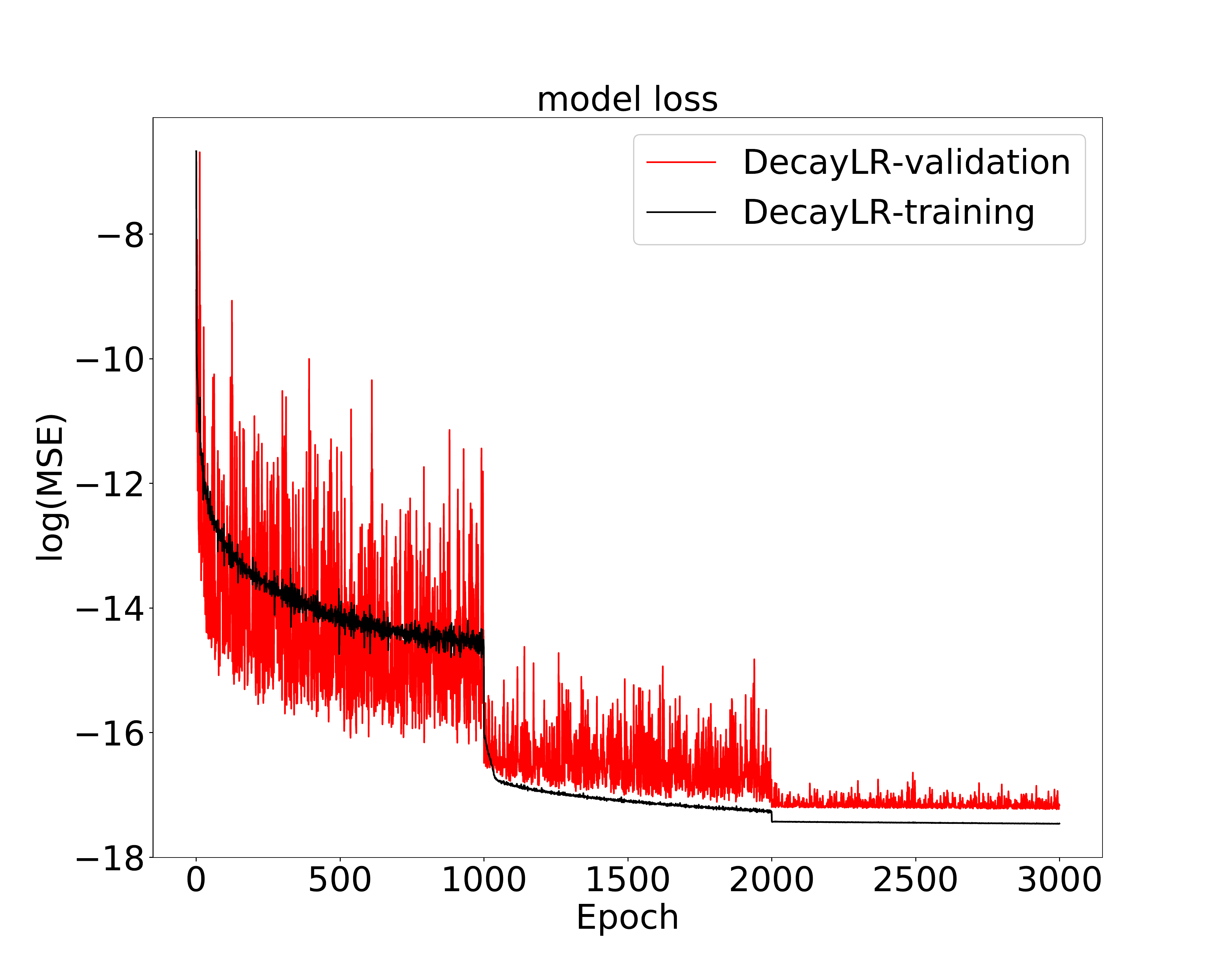} }  
      \subfloat[Losses with cyclical learning rates] {\label{fig:CLR_train_valuation_loss_heston}     
        \includegraphics[width=8cm]{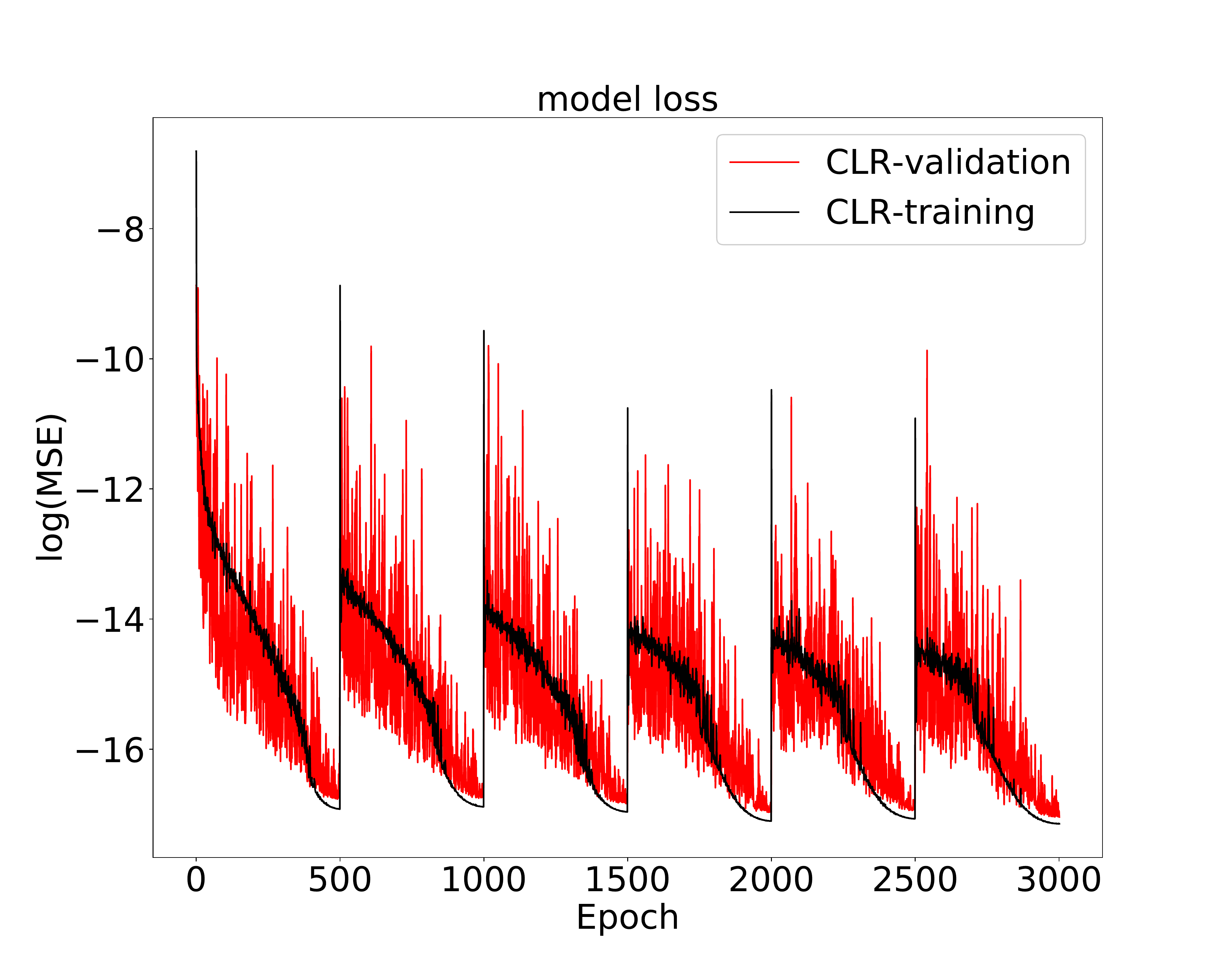} }
    \caption{The history of training and validation losses for the Heston model.} 
    \label{fig:train_valuation_loss_heston}
\end{figure}

We present, as an example, the results of the training stage of the ANN solver for the Heston model option prices to compare three different learning rate schedules. Figure \ref{fig:train_valuation_loss_heston} demonstrates that the training error and the validation error agree well and that over-fitting does not occur when using these schedules. As shown in Figure \ref{fig:nn_training_loss_comparison}, in this case a decay rate based schedule outperforms the CLR with the same learning rate bounds, although with the CLR the difference between training and validation losses are smaller.  This is contrary to the conclusion in~\cite{SmithCyclicalNetworks}, but their network included batch normalization and L2 regularization. For the tests in this paper, we will employ the CLR  to find the optimal range of learning rates, which is then applied in the Decay\-LR  schedule to train the ANNs. 

\section{Numerical Results} \label{section:Numerical Results}

We show the performance of the ANNs for solving the financial models, based on the following  accuracy metrics (which forms the basis for the training),

\begin{equation}
\text{MSE} = \frac{1}{n} \sum (y_i-\hat{y}_i)^2,
\end{equation}
where $y_i$ is the actual value and $\hat{y}_i$ is the ANN predicted value. The MSE is used as the training metric to update the weights, and all above metrics are employed to evaluate the selected ANN.  
For completeness, however, we also report the other well-known metrics,

\begin{subequations}
\begin{align}
	\text{RMSE} &= \sqrt{\text{MSE}}, \\
\text{MAE} &= \frac{1}{n} \sum |y_i-\hat{y}_i|, \\
\text{MAPE} &= \frac{1}{n} \sum \frac{|y_i-\hat{y}_i|}{y_i}. 
\end{align}
\end{subequations}

We start with the Black-Scholes model which gives us closed-form option prices, that are learned by the ANN. We also train the ANN to learn the implied volatility, based on the iterative root-finding Brent method. Finally, the ANN learns the results obtained by the COS method to solve the Heston model with a number of different parameters.   

\subsection{Details of the data set} \label{section: data set impact}

As a data-driven approach, the quality of a data set has an impact on the performance of the resulting model. Theoretically, an arbitrary number of samples can be generated since the mathematical model is known. In reality, a sampling technique with good space-filling properties should be preferable.  Latin hypercube sampling (LHS) \cite{McKay1979ACode} is able to generate random samples of the parameter values from a multidimensional distribution, resulting in a better representation of the parameter space. When the sample data set for the input parameters is available, we select the appropriate  numerical methods to generate the training results.  For the Black-Scholes model, the option prices are obtained by the closed-form formula. For the Heston model, the prices are calculated by the COS method with a robust COS method version. With the Heston prices determined, Brent's method will be used to find the corresponding implied volatility.  The whole data set is randomly divided into two groups, 90\% will be the training  and  10\% the test set.

\begin{table}[htp]
\begin{center}
\caption{The different sizes of training data set when training the ANN}
  \begin{tabular}{  c | c | c |c |c |c |c |c }
    \hline
    Case &  0 & 1 & 2 & 3 & 4& 5& 6 \\ \hline
    Training size ($\times 24300$) &  1/8 & 1/4 & 1/2 & 1 & 2 & 4 & 8 \\
    \hline	
  \end{tabular}
 \label{table:The training size.}
 \end{center}
  \end{table} 

\begin{figure}[htp]
\includegraphics[width = 1.0\textwidth]{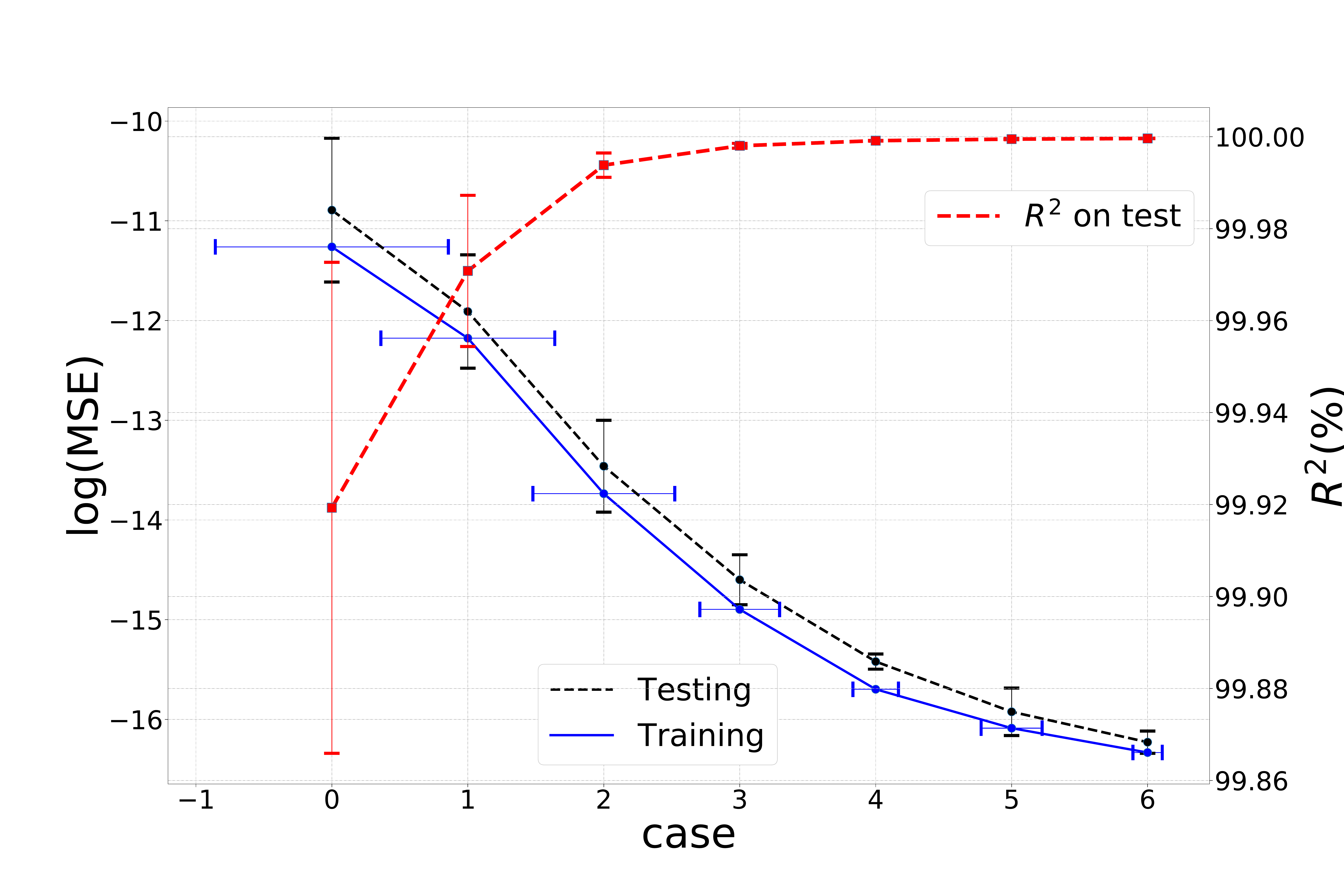}
\caption{$R^2$ and MSE vs. size of the training set. The figure shows improvement in the loss, which is an
indicator of the model performance.}
\label{fig:r2_mse_against_sample_size_randomsample}
\end{figure}

In order to investigate the relation between the prediction accuracy and  the size of the training set, we increase the number of training samples from $\frac{1}{8}$ to 8 times the baseline set. Meanwhile, the test data is kept unchanged. The example here is learning the implied volatility. We first train the ANN for each data set by using a decaying learning rate, as described in Section \ref{section:learning_rate}, and repeat the training stage for each case 5 times with different random seeds and average the model performance. As shown in Figure \ref{fig:r2_mse_against_sample_size_randomsample}, with an increasing data size, the prediction accuracy increases and the corresponding variance, indicated by the error bar, decreases. So, we employ random search for the hyper-parameters on a small-sized data set, and train the selected ANN on a large data set. 
The schedule of decaying learning rates is as discussed in Section \ref{section:learning_rate}. The training and validation losses remain close, which indicates that there is no over-fitting.

\subsection{Black-Scholes model} \label{sec:BS-result}

Focusing on European call options, we generate 1,000,000 random samples for the input parameters, see Table \ref{table:black-scholes para range}. We calculate the corresponding European option prices $V(S,t)$ of Equation (\ref{eq:bs-pde}) with the solution in~(\ref{eq:bs_solution}). As a result, each sample contains five variables \{$S_0/K, \tau, r, \sigma, V/K$\}.  The training samples are fed into the ANN, where the input includes \{$S_0/K, \tau,r, \sigma$\}, and the output is the scaled option price $V/K$.

\begin{table}[htp]
\begin{center}
\caption{Wide and narrow Black-Scholes parameter ranges}
  \begin{tabular}{ c| c | c | c | c }
    \hline

     & Parameters                  &  Wide Range & Narrow Range & Unit  \\ \hline

\multirow{4}{*}{Input} & Stock price($S_0/K$) & [0.4, 1.6]  & [0.5, 1.5]  & - \\

    & Time to Maturity($\tau$)    & [0.2, 1.1] & [0.3, 0.95] & year \\

    & Risk free rate($r$)         & [0.02,  0.1] & [0.03,  0.08] & -\\

    & Volatility($\sigma$)        & [0.01, 1.0] & [0.02, 0.9] &  - \\

    \hline

    Output & Call price($V/K$)           & (0.0, 0.9) & (0.0, 0.73) & -\\

    \hline	  

  \end{tabular}
  
 \label{table:black-scholes para range}
 \end{center}
  \end{table}

We distinguish during the evaluation of the ANN, two different test data sets, i.e. a wide test set and a slightly more narrow test set.
The reason is that we observe that often the ANN approximations in the areas very close to parameter domain boundaries may give rise to somewhat larger approximation errors,  and the predicted values in the middle part are of higher accuracy. We wish to alleviate this boundary-related issue. 

\begin{figure}[H]
    \centering
    \subfloat[Error distribution (wide set)] {\label{fig:bs_error_distr}
        \includegraphics[width=8cm]{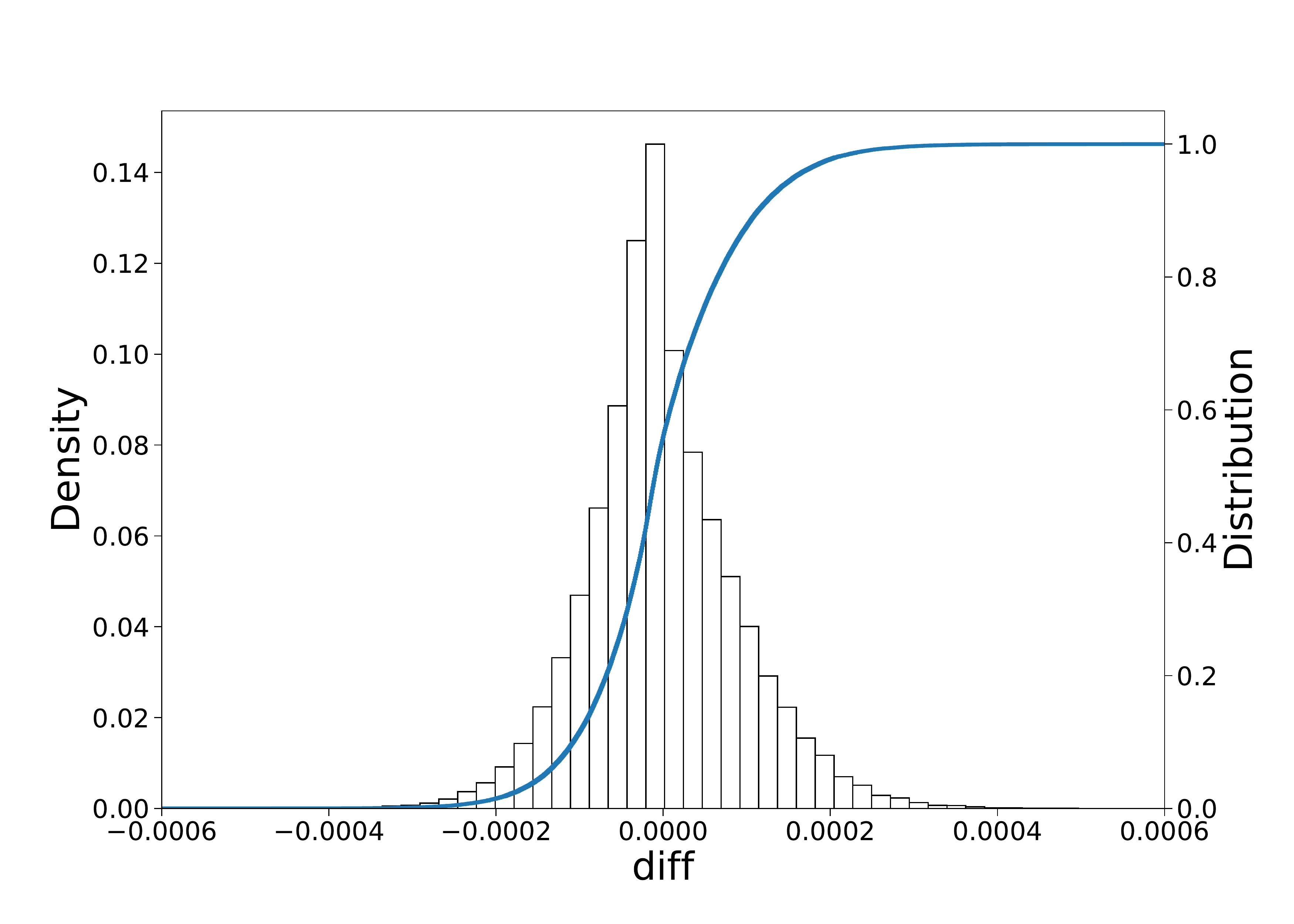} } 
 \subfloat[Error distribution (narrow set)] {\label{fig:bs_error_distr narrow}     
        \includegraphics[width=8cm]{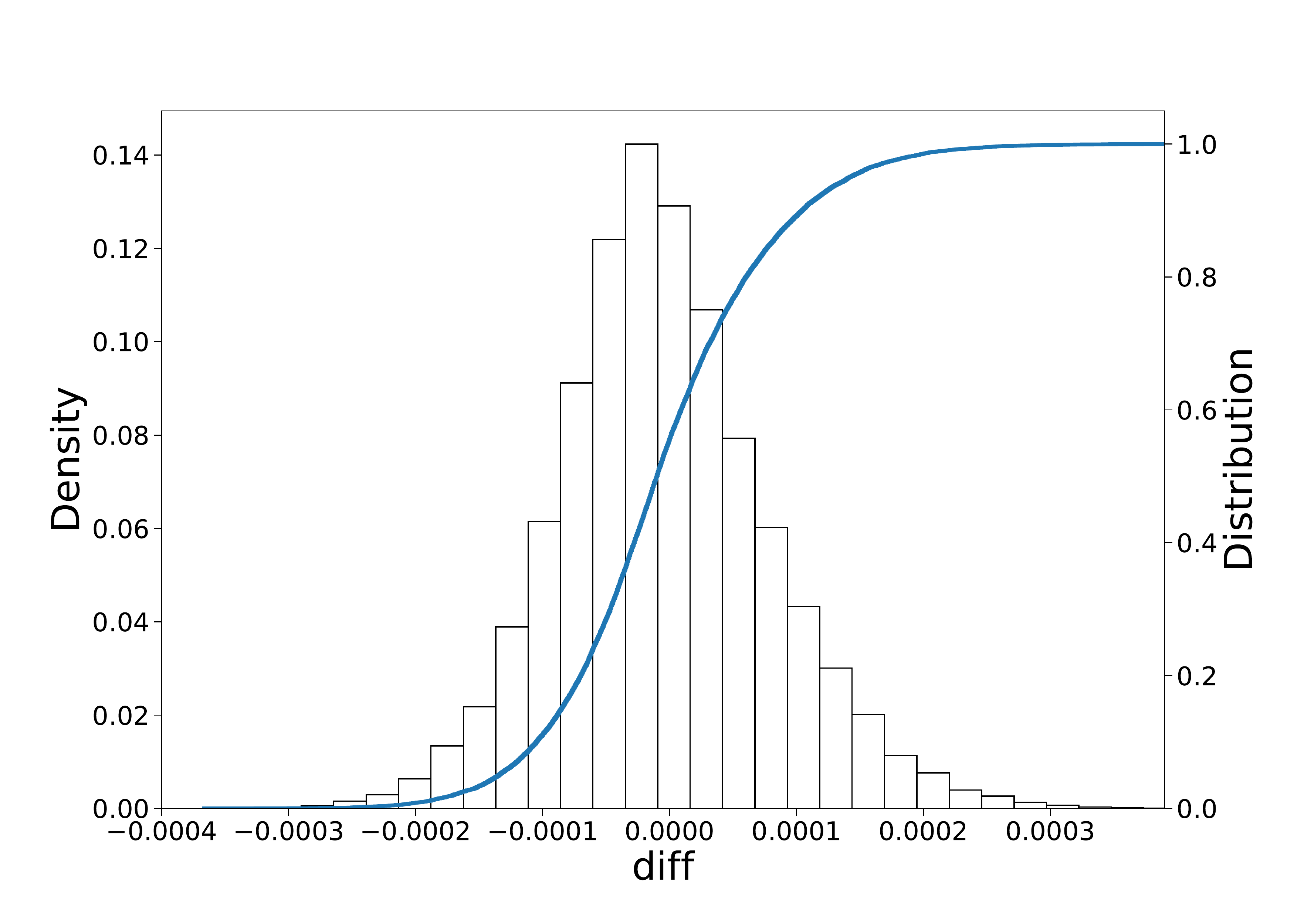} }
    \caption{Left: ANN performance on the wide test data set. Right: ANN performance on the narrow test data set.}    
    \label{fig:bs_nn_test}
\end{figure}

The wide test data set is based on the same parameter ranges as the training data set. As shown in Table \ref{table:NN-BS-Performance}, the root averaged mean-squared error (RMSE) is around $9\cdot 10^{-5}$, which is an indication that the average pricing error is 0.009\% of the strike price. Figure \ref{fig:bs_error_distr} shows the histogram of prediction errors, where it can be seen that the error approximately exhibits a normal distribution, and the maximum absolute error is around 0.06\%.

The narrow test set is based on a somewhat more narrow parameter range than the training data set. As Table \ref{table:NN-BS-Performance} shows,  when the range of parameters in the test set is smaller than the training data set, ANN's test performance slightly improves. Figure \ref{fig:bs_nn_test} shows that the largest deviation becomes smaller, being less than 0.04\%. The goodness of fit $R^2$-criterion measures the distance between the actual values and the predicted ones. There is no significant difference in $R^2$ in both cases.

Overall, it seems a good practice to train the ANN on a (slightly too) wide data set, when the parameter range of interest is somewhat smaller. In the next sections, however, we list the performance of the ANN on the wide test data set.


\begin{table}[H]
\begin{center}
\caption{BS-ANN performance on the test data set}
\scalebox{1.0}{

  \begin{tabular}{  c | c | c | c | c }
    \hline

          BS-ANN               &  MSE & RMSE & MAE & MAPE    \\ \hline

     Training-wide & $8.04\cdot 10^{-9}$ & $8.97\cdot 10^{-5}$& $6.73\cdot 10^{-5}$ & $3.75\cdot 10^{-4}$ \\

     Testing-wide  & $8.21\cdot 10^{-9}$& $9.06\cdot 10^{-5}$ & $6.79\cdot 10^{-5}$&  $3.79\cdot 10^{-4}$ \\

     Testing-narrow  & $7.00\cdot 10^{-9}$& $8.37\cdot 10^{-5}$ & $6.49\cdot 10^{-5}$&  $3.75\cdot 10^{-4}$ \\
    \hline
  \end{tabular} }
  
 \label{table:NN-BS-Performance}
 \end{center}
  \end{table}

\subsection{Implied volatility}

The aim here is to learn the implicit relationship between implied volatilities and option prices, which is guided by Equation (\ref{eq:ivf}). The option Vega can become arbitrarily small, which may give rise to a steep gradient problem in the ANN context.  It is well-known that an ANN may generate significant prediction errors in regions with large gradients. We therefore propose a gradient-squash approach to handle this issue.

First of all, each option price can be split into a so-called {\em intrinsic value} and a {\em time value}, and we subtract the intrinsic value, as follows,

$$\tilde{V} = V_t - \max(S_t-Ke^{-r\tau},0),$$
where $\tilde{V}$ is the option time value. Note that this change only applies to ITM options, since the OTM intrinsic option value is equal to zero. 
The proposed approach to overcome approximation issues is to reduce the gradient's steepness by furthermore working under a log-transformation of the option value. The resulting input is then given by $\{\log{(\tilde{V}/K)}, S_0/K, r, \tau\}$.  The adapted gradient approach increases the prediction accuracy significantly.

\subsubsection{Model performance} 

In this case, the data samples can be created in a forward stage, i.e.,  we will work with the Black-Scholes solution (instead of the root-finding method) to generate the training data set. 
Given $\sigma$, $\tau$, $K$, $r$ and $S$, the generator, i.e. the Black-Scholes formula, gives us the option price $V(t_0,S_0)=BS(S_0, K, \tau, r, \sigma)$. For the data collection $\{V, S_0, K, \tau,r,\sigma\}$, we then take the input $\sigma$ as the implied volatility $\sigma^*\equiv \sigma$ and place it as the output of the ANN.  Meanwhile, the other variables $\{V, S_0, K, \tau, r\}$  will become the input of the ANN, followed by the log-transformation $\log(\tilde{V}/K)$. In addition, we do not take into consideration the samples whose time values are extremely small, like those for which $\tilde{V}<10^{-7}$.

\begin{table}[htp]
\begin{center}
 \caption{ Parameter range of data set}
  \begin{tabular}{ c| c | c | c }
    \hline
     & Parameters                  &  Range & Unit  \\ \hline

    \multirow{4}{*}{NN Input} & Stock price ($S_0/K$)          & [0.5, 1.4]  & - \\

    & Time to maturity ($\tau$)    & [0.05, 1.0] & year \\

    & Risk-free rate ($r$)         & [0.0,  0.1] & -\\    

    & Scaled time value ($\log{(\tilde{V}/K)}$)           & [-16.12, -0.94] & -\\

    \hline
    NN output & Volatility ($\sigma$)        & (0.05, 1.0)&  - \\
    \hline	 
 \end{tabular}
 \label{table:iv para range}
 \end{center}
 \end{table}

\begin{figure}[H]
    \centering
    \subfloat[Comparison of implied volatilities] {\label{fig:iv_compare_pred_actual_test}
        \includegraphics[width=8cm]{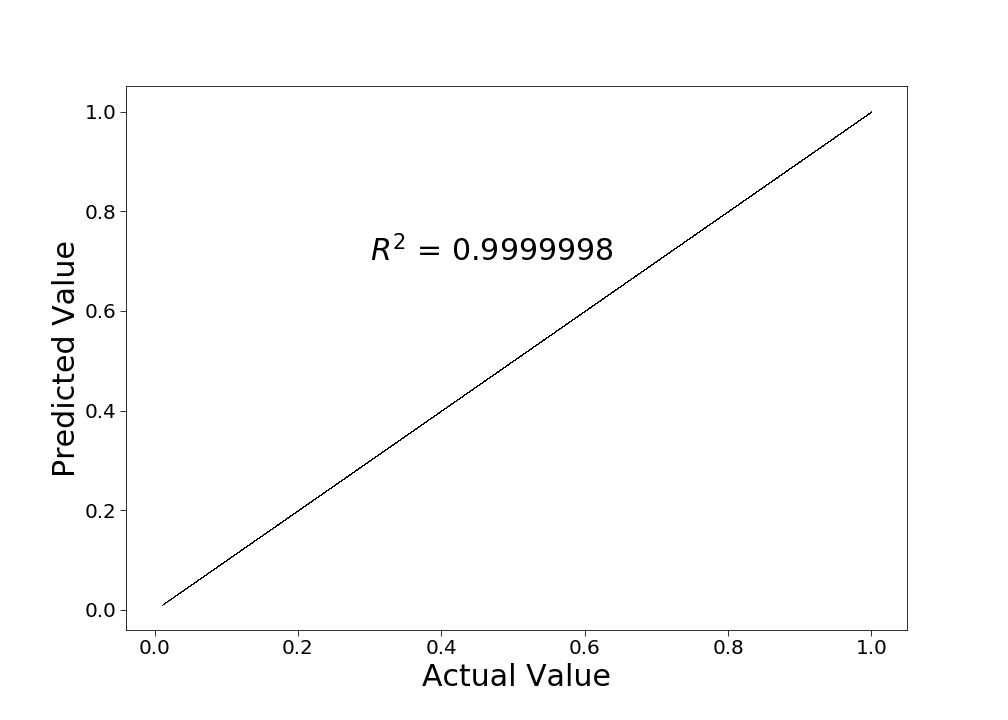} }  
      \subfloat[The error distribution] {\label{fig:iv_error_distr_test}    
        \includegraphics[width=8cm]{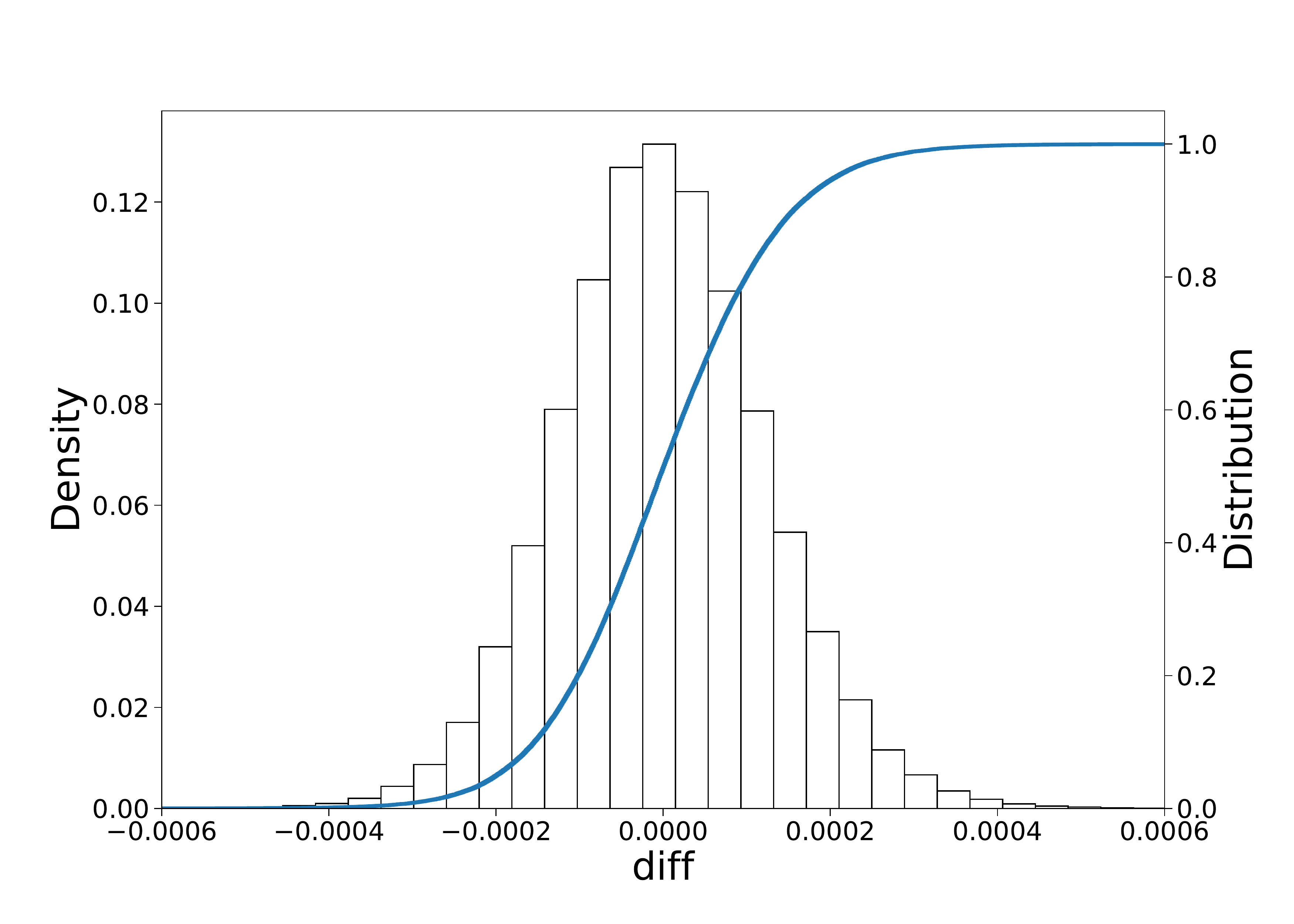} }
    \caption{Out-of-Sample IV-ANN performance on the scaled input. }
    \label{fig:iv_nn_test}
\end{figure}

Table \ref{table:perfomance_iv_NN_scale} compares the performance of the ANN with the scaled and original (unscaled) input, where it is clear that scaling improves the ANN performance significantly.  Figure \ref{fig:iv_nn_test} shows the out-of-sample performance of the trained ANN on the scaled inputs. The error distribution also approximately follows a normal distribution, where the maximum deviation is around $6\cdot 10^{-4}$, and most of implied volatilities equal their true values.

 \begin{table}[H] 
\begin{center}
\caption{Out-of-Sample ANN performance comparison}
\scalebox{1.0}{
  \begin{tabular}{  c | c | c | c | c}
    \hline
                             &  MSE & MAE & MAPE & $R^2$  \\ \hline

     \makecell{ Input: $m$, $\tau$, $r$, $V/K$ \\ Output: $\sigma^*$}  & 6.36 $\cdot 10^{-4}$ & 1.24 $\cdot 10^{-2}$  & 7.67 $ \cdot 10^{-2}$ &  0.97510 \\
     
\hline

    \makecell{Input: $m$, $\tau$, $r$, $log(\tilde{V}/K)$ \\Output: $\sigma^*$}  & 1.55 $\cdot 10^{-8}$  &  9.73 $\cdot 10^{-5}$  & 2.11 $\cdot 10^{-3}$  & 0.9999998\\
    \hline
  \end{tabular} }
 
 \label{table:perfomance_iv_NN_scale}
 \end{center}
  \end{table}

\subsubsection{Comparison of root-finding methods}

We compare the  performance of five different implied-volatility-finding methods, including IV-ANN, Newton-Raphson, Brent, the secant and the bisection method, in terms of run-time on a CPU and on a GPU.
For this purpose, we compute 20,000 European call options for which all numerical methods can find the implied volatility. The $\sigma$-value range for bisection and for Brent's method is set to $[0, 1.1]$, and the initial guess for the Newton-Raphson and secant method equals  $\sigma_0^*=0.5$. The true volatility varies in the range $[0.01, 0.99]$, with the parameters, $r$=0, $T$= 0.5, $K$=1.0, $S_0$=1.0.

Table \ref{table:iv-Performance-comparison} shows that Brent's method is the fastest {\em among the robust} iterative methods (without requiring domain knowledge to select a suitable initial value). From a statistical point-of-view, the ANN solver gives rise to  an acceptable averaged error  MAE $\approx10^{-4}$, and, importantly, its computation is faster by a factor 100 on a GPU and 10 on a CPU, as compared to the Newton-Raphson iteration. By the GPU architecture, the ANN processes the input 'in batch mode', calculating a number of implied volatilities simultaneously, which is the reason for the much higher speed. Besides, the acceleration on the CPU is also obvious, as only matrix multiplications or inner products are required.

  \begin{table}[H]
\begin{center}
  \caption{Performance comparison: CPU (Intel i5, 3.33GHz with cache size
4MB) and GPU(NVIDIA Tesla P100) }
  \begin{tabular}{  c | c | c | c }
    \hline

    Method                  &  GPU (sec) & CPU (sec) & Robustness \\ \hline

    Newton-Raphson          & 19.68 & 23.06   & No  \\

    Brent                   & 52.08 &  60.67  & Yes \\ 

    Secant 		            & 88.73 & 103.76  & No   \\ 

    Bi-section              & 337.94 &  390.91 &Yes \\

    \hline

    IV-ANN         & 0.20 &  1.90  & Yes   \\

    \hline
  \end{tabular}

 \label{table:iv-Performance-comparison}
 \end{center}
  \end{table}

\subsection{The Heston stochastic volatility model}


This section presents the quality of the ANN predictions of the Heston option prices and the corresponding implied volatilities. The performance of the Heston-ANN solver is also evaluated. 

\subsubsection{Heston model for option prices}

The Heston option prices are computed by means of the COS method in this section. The solution to the Heston model also can be obtained by other numerical techniques, like PDEs discretization or Monte Carlo methods.   The COS method has been proved to guarantee a high accuracy  with less computational expense. 

According to the given ranges of Heston parameters in Table \ref{table:heston para-range-random-vol}, for the COS method, the integration interval is based on $L_{COS}=50$, with the number of Fourier cosine terms in the expansion being $N_{COS}=1500$. 
The prices of deep OTM European call options are calculated using the put-call parity, as the COS method call prices that are close to zero may be inaccurate due to truncation errors. In Table \ref{table:heston para-range-random-vol}, we list the range of the six Heston input parameters ($r$, $\rho$, $\kappa$, $\bar{\nu}$, $\gamma$, $\nu_0$) as well as the two option contract-related parameters ($\tau$, $m$), with a fixed strike price, $K$=1. We generate around one million data points by means of the Latin hypercube sampling, using 10\% as testing, 10\% as validation and 80\% as the training data set. After 3,000 epochs with a decaying learning rate schedule, as shown in Table \ref{table:ANN-Heston Performance}, the Heston-ANN solver has been well trained, avoiding over-fitting and approximating the prices accurately.  Although the number of input parameters is doubled as compared to the Black-Scholes model, the Heston-ANN accuracy is also highly satisfactory and the error pattern is similar to that of the BS-ANN solver. 

   \begin{table}[H]
\begin{center}
  \caption{The Heston parameter ranges for traing the ANN}
  \begin{tabular}{  c|c | c | c }
    \hline

    ANN &Parameters                  &  Range     & Method  \\ \hline

    \multirow{9}{*}{NN Input} & Moneyness, $m=S_0/K$          & (0.6, 1.4)  & LHS \\ 

    &Time to maturity, $\tau$       & (0.1, 1.4)(year) & LHS \\

    &Risk free rate, $r$         & (0.0\%, 10\%) & LHS \\

    &Correlation, $\rho$         &  (-0.95, 0.0) & LHS \\

    &Reversion speed, $\kappa$   & (0.0, 2.0) & LHS \\

    &Long average variance, $\bar{\nu}$  & (0.0, 0.5) & LHS\\

    &Volatility  of volatility, $\gamma$  & (0.0, 0.5) & LHS \\   

    &Initial variance, $\nu_0$    & (0.05, 0.5)  & LHS\\  

    \hline

    NN output& European call price, $V$    &  (0,  0.67) & COS\\    

    \hline	  

  \end{tabular}
 \label{table:heston para-range-random-vol}
 \end{center}
  \end{table}

\begin{table}[H]
\caption{ The trained Heston-ANN performance }
\begin{center}
\scalebox{1.0}{
  \begin{tabular}{  c | c | c | c | c   }
    \hline

          Heston-ANN   & MSE  &  MAE & MAPE & $R^2$  \\ \hline
     Training &   $1.34\cdot 10^{-8}$   & $ 8.92\cdot 10^{-5}$& $ 5.66\cdot 10^{-4}$ & 0.9999994 \\
     Testing &    $1.65\cdot 10^{-8}$  & $9.51\cdot 10^{-5}$ & $6.27\cdot 10^{-4}$ &  0.9999993 \\
    \hline	  
\end{tabular} }

\label{table:ANN-Heston Performance}
\end{center}
\end{table}

 
\begin{figure}[H]
    \centering
    \subfloat[COS vs. Heston-ANN prices] {\label{fig:heston_compare_price}
        \includegraphics[width=8cm]{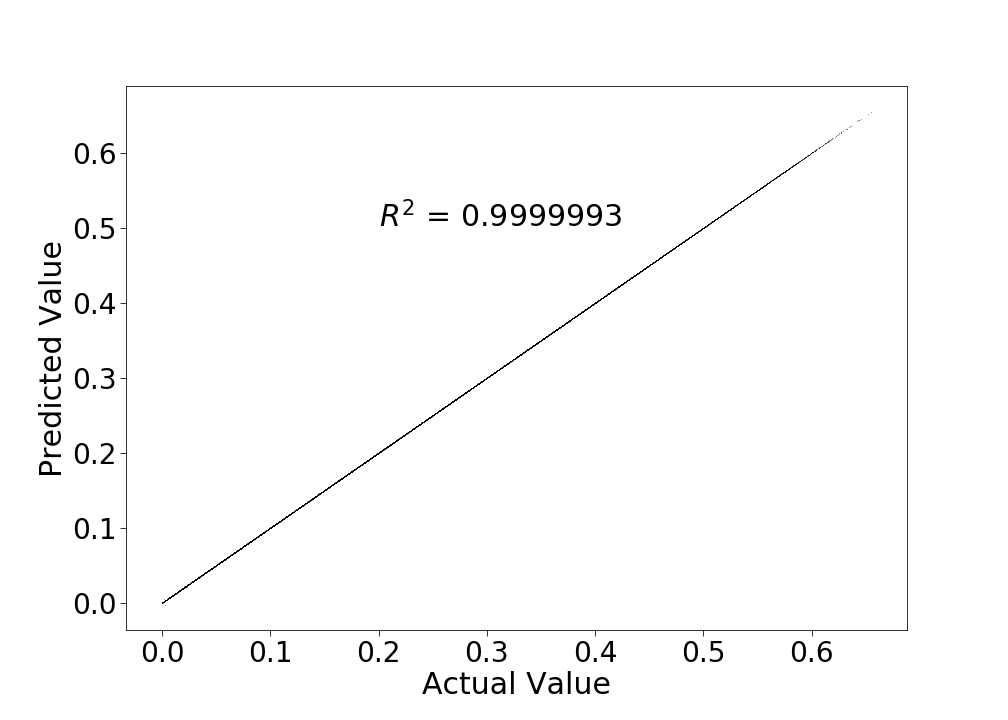} }  
      \subfloat[The error distribution] {\label{fig:heston_error_distr}   
        \includegraphics[width=8cm]{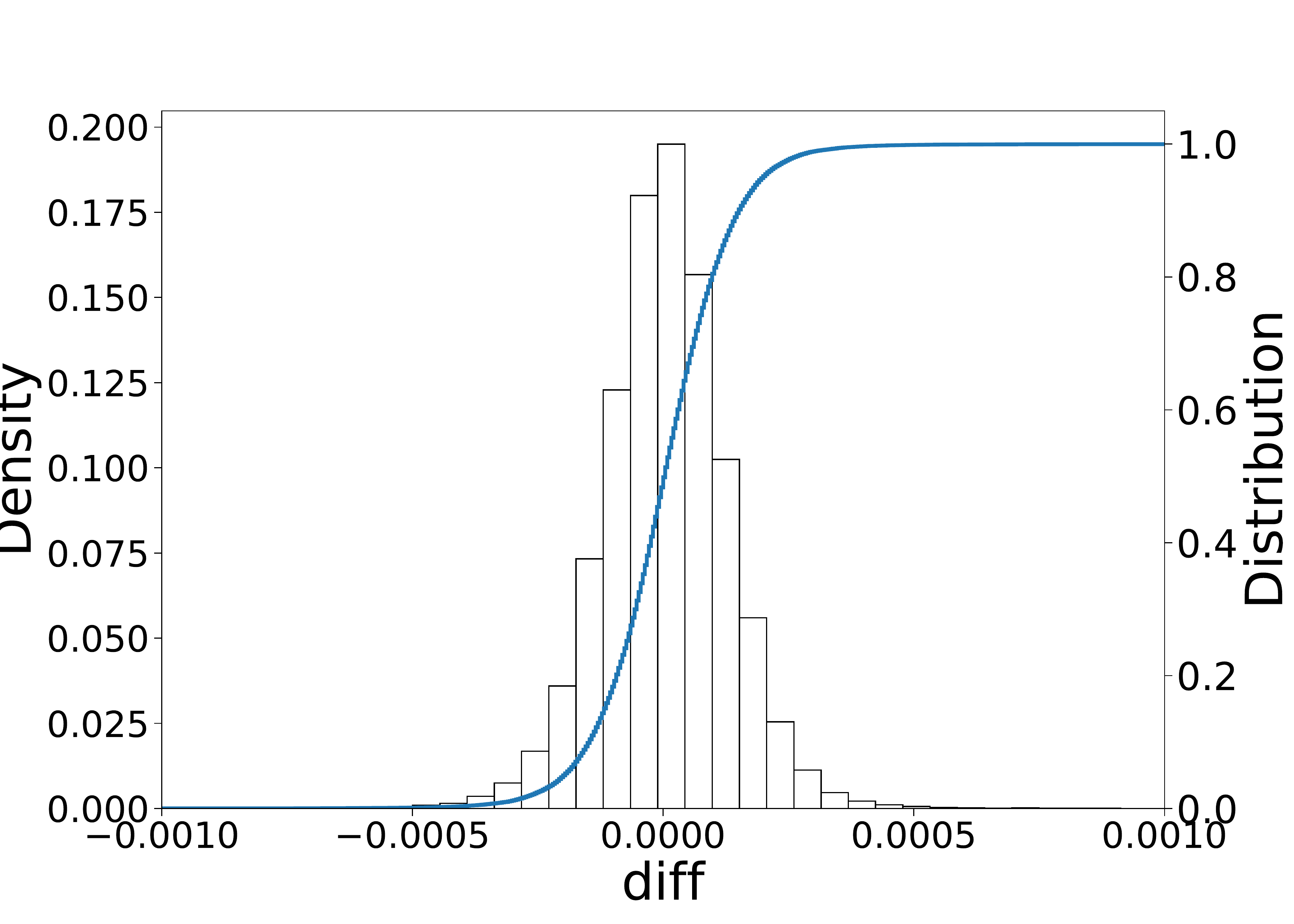} }
    \caption{ Out-of-sample Heston-ANN performance.  }
    \label{fig:heston_nn_test}
\end{figure}

\subsubsection{Heston model and implied volatility} \label{section:heston-iv}

We design two experiments to illustrate the ANN's ability of computing the implied volatility based on the Heston option prices. In the first experiment, the ground truth for the implied volatility is generated by means of two steps. Given the Heston input parameters, we first use the COS method to compute the option prices, after which we use Brent's method to compute the Black-Scholes implied volatility $\sigma^\ast$. The machine learning approach is also based on two steps. First of all, the Heston-ANN is used to compute the option prices, and, subsequently, we use IV-ANN to compute the corresponding implied volatilities. We compare these two approaches as follows. 

\begin{figure}[H]
\begin{center}
\includegraphics[width = 1.0\textwidth]{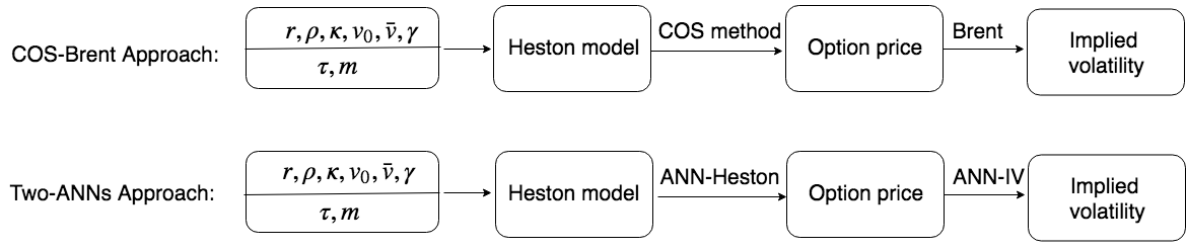}
\caption{ Two approaches of computing implied volatility for Heston model.}
\label{fig:ann_heston_iv_flowchart}
\end{center}
\end{figure}

\begin{figure}[H]
    \centering
    \subfloat[Case 1: The error distribution] {\label{fig:heston_nn2nn_error_iv}
        \includegraphics[width=8cm]{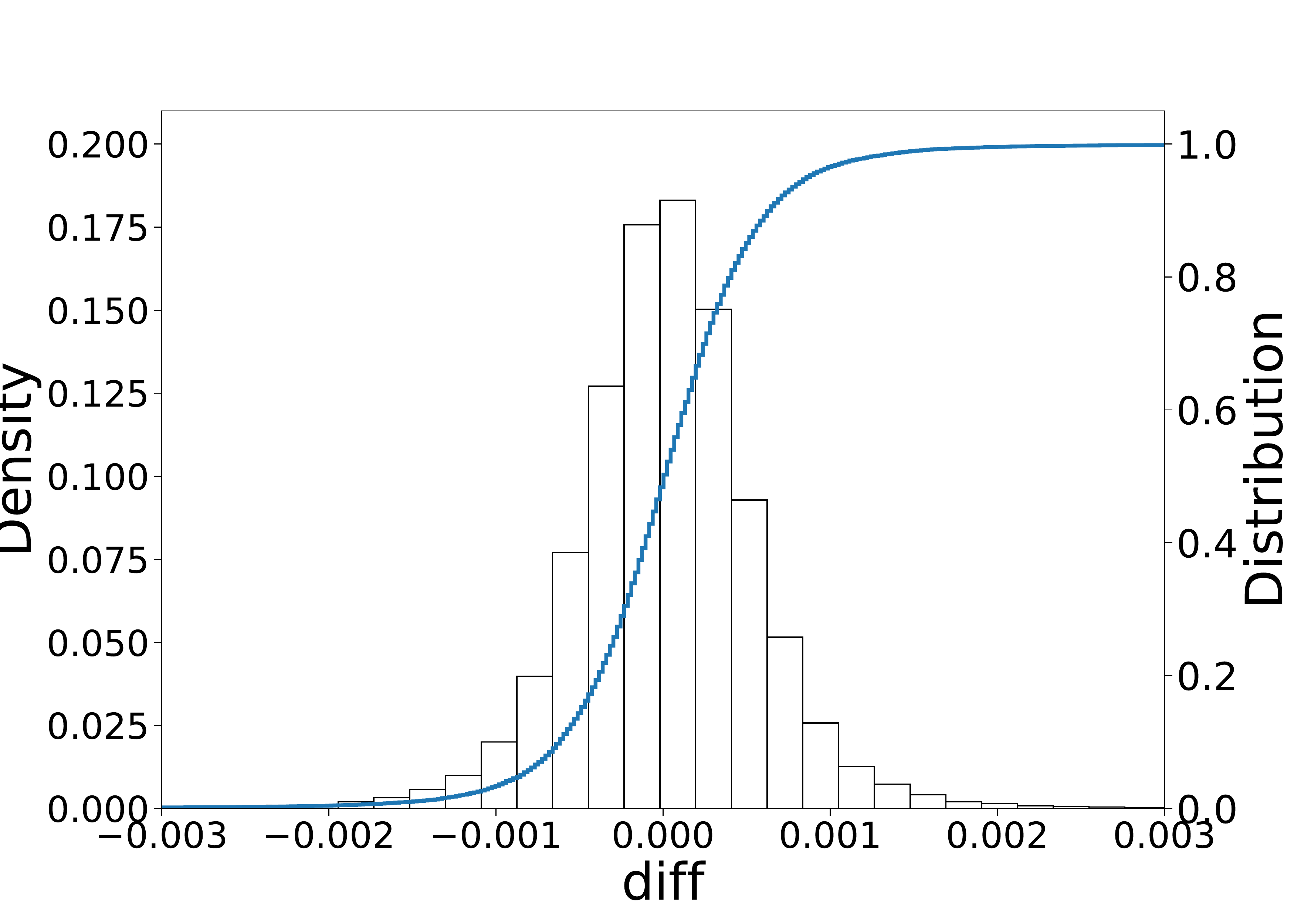} }  
      \subfloat[Case 2: The error distribution] {\label{fig:heston_nn2nn_error_iv_narrower}    
        \includegraphics[width=8cm]{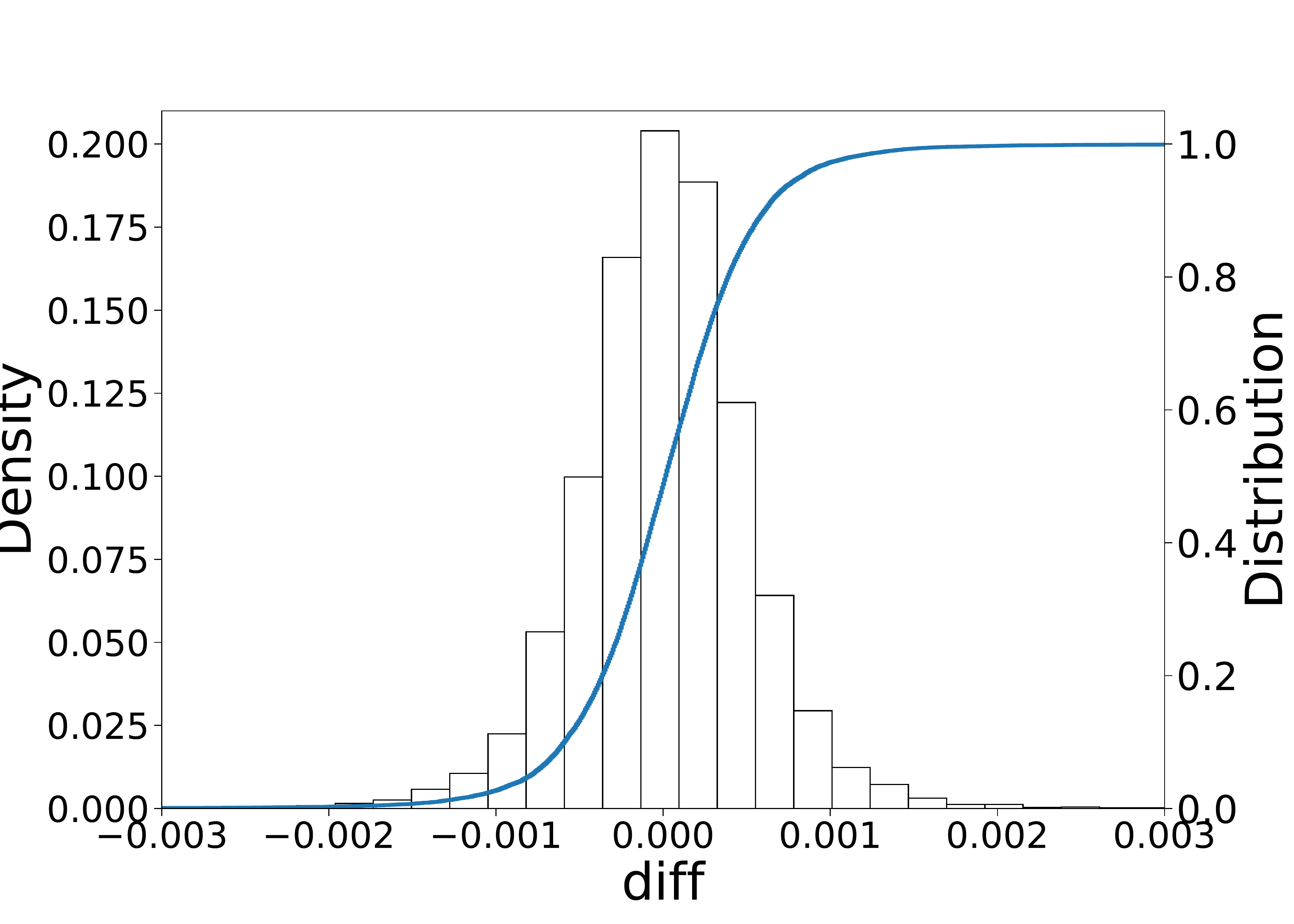} }
\caption{The error distribution of the implied volatility: The combined Heston-ANN and IV-ANN  techniques for implied volatility.} 
\label{fig:dnn_heston_iv_perf_narrower}
\end{figure}

Note that the ANN solver performs best in the mid of all parameter ranges, and tends to become worse at the domain boundaries.  We therefore first choose the range of moneyness $m \in [0.7, 1.3]$ and the time to maturity $\tau \in [0.3, 1.1]$. Table \ref{table:ANN-Heston-IV Performance} shows the overall performance of the ANN. As the IV-ANN takes the output of the Heston-ANN as the input, the accumulated error reduces the overall accuracy slightly. However, the root averaged mean error is still small, RMSE$\approx 7\cdot 10^{-4}$. Then we  reduce the range of the parameters, as listed in the third row of Table \ref{table:ANN-Heston-IV Performance}, and find that the prediction accuracy increases with the parameter ranges shrinking.  Comparing the results in Figure \ref{fig:dnn_heston_iv_perf_narrower} and Table \ref{table:ANN-Heston-IV Performance}, the goodness of fit as well as the error distribution improve  with the slightly smaller parameter range, which is similar to our findings for the BS-ANN solver.


\begin{table}[htp]
\begin{center}
\caption{ Out-of-sample performance of the Heston-ANN plus the IV-ANN }
\scalebox{1.0}{
  \begin{tabular}{  c | c | c | c | c }
    \hline
    Heston-ANN \& IV-ANN   & RMSE & MAE & MAPE & $R^2$  \\ \hline  
     \makecell{Case 1:  \\ $\tau\in[0.3, 1.1]$, $m\in[0.7, 1.3] $ }   & 7.12 $\cdot 10^{-4}$ & 4.19 $\cdot 10^{-4}$ &1.46 $\cdot 10^{-3}$&  0.999966 \\
     \hline
     \makecell{Case 2: \\ $\tau\in[0.4, 1.0]$, $m\in[0.75, 1.25] $ } 
     & 5.53 $\cdot 10^{-4}$& 3.89 $\cdot 10^{-4}$ &1.14 $\cdot 10^{-3}$&  0.999980 \\
     
    \hline	  
\end{tabular} }

\label{table:ANN-Heston-IV Performance}
\end{center}
\end{table}

\begin{figure}[H]
    \centering
    \subfloat[Implied volatility surface] {\label{fig:heston_ivol_surface dnn}
        \includegraphics[width=7cm]{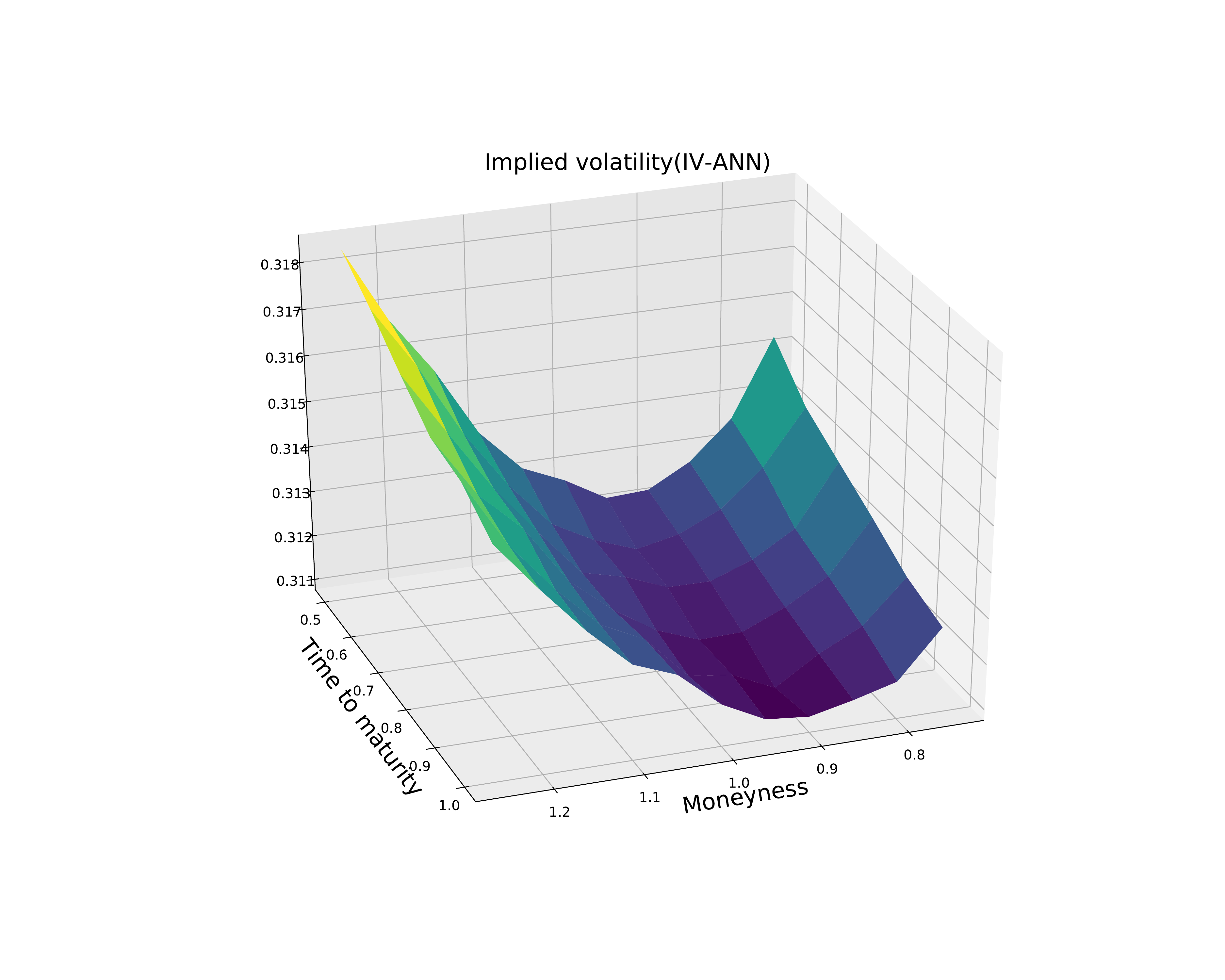} } 
      \subfloat[Implied volatility difference] {\label{fig:heston_ivol_surface brent}   
        \includegraphics[width=7cm]{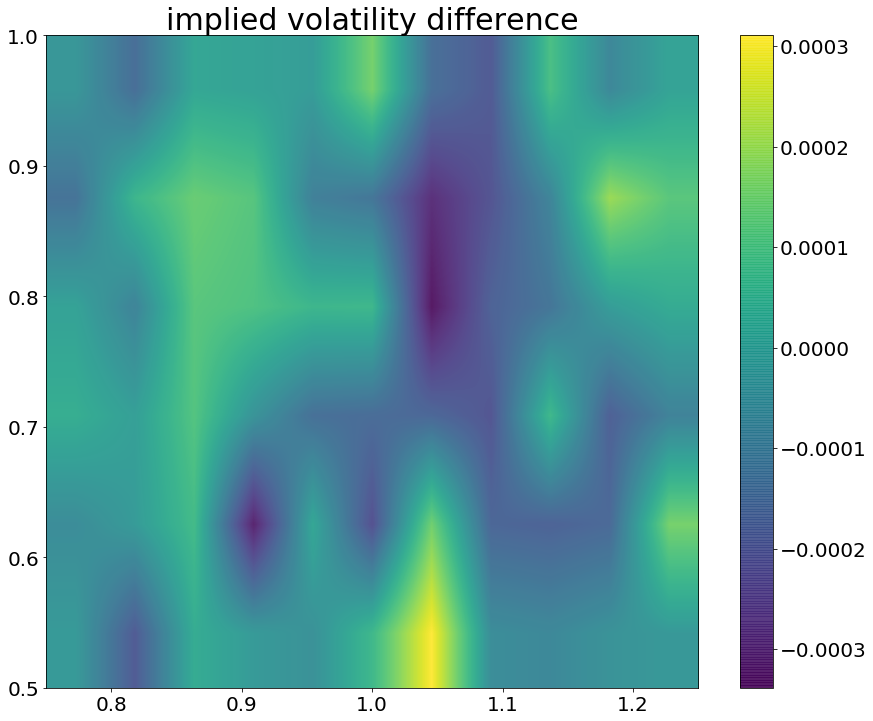} }
 \caption{ 
Left: Heston implied volatility surface generated by IV-ANN.
Right: Heston implied volatility difference between Brent's method and IV-ANN. } 
 \label{fig:one_ivol_surface}
\end{figure}

Another experiment is to show that the IV-ANN can generate a complete implied volatility surface for the Heston model. With the following Heston parameters (an example to produce a smile surface), $\rho$=-0.05, $\kappa$=1.5, $\gamma$=0.3, $\bar{v}$=0.1, $v_0$=0.1 and $r$=0.02, we calculate the option prices by the COS method for the moneyness $m \in [0.7, 1.3]$ and time to maturity $\tau \in [0.5, 1.0]$. The implied volatility approximated by means of the IV-ANN is shown in Figure \ref{fig:heston_ivol_surface dnn}, and the maximum deviation between the ground-truth and the predicted values is no more than than 4$\cdot 10^{-4}$.

Concluding, the ANN can approximate the Heston option prices as well as the implied volatilities accurately. The characteristic function of the financial model is not required during the test phase of the ANN.

\section{Conclusions and discussion}
In this paper we have proposed an ANN approach to reduce the computing time of pricing financial options, especially for high-dimensional financial models.  We test the ANN approach on three different solvers, including the closed-form solution for the Black-Scholes equation, the COS method for the Heston model and Brent's root-finding method for the implied volatilities. Our numerical results show that the ANN can compute option prices and implied volatilities efficiently and accurately in a robust way. This means, particularly for asset price processes leading to much more time-consuming computations, that  we are able to provide a highly efficient approximation technique by means of the ANN. Although the off-line training will take longer then, the on-line prediction will be fast. Moreover, parallel computing allows the ANN solver to process derivative contracts ``in batch mode'' (i.e. dealing with many observed market option prices simultaneously during calibration), and this property boosts the computational speed by a factor of around 100 on GPUs over the original solver in the present case.   
We have shown that the boundaries of parameter values have an impact  when applying the ANN solver. It is recommended to train the ANN on a data set with somewhat wider ranges than values of interest.
Regarding high-dimensional asset models,  as long as the option values can be obtained by any numerical solver (Fourier technique, finite differences or Monte Carlo method), we may speed up the calculation by employing a trained ANN.

Although we focus on European call options in this work,  it should be possible to extend the approach to  pricing more complex options, like American, Bermuda or exotic options.  This  work initially demonstrates the feasibility of learning a data-driven solver to speed up solving parametric financial models.  The model accuracy can be further improved, for example, by using deeper neural networks, more complex NN architectures. The solver's speed may also improve, for example, by designing a more shallow neural network, extracting insight from the complex network~\cite{Hinton2015DistillingNetwork}. 

Furthermore, the option Greeks, representing the sensitivity of option prices with respect to the market or model parameters, are important in practice (i.e. for hedging purposes). As ANNs approximate the solution to the financial PDEs, the related derivatives can also be recovered from the trained ANN. There are several ways to calculate Greeks from the ANN solver. A straightforward way is to extract the gradient information directly from the ANN, since the approximation function in Equation (\ref{eq:nn-formula}) is known analytically. Alternatively, a trained ANN may be interpreted as an implicit function, where Auto-Differentiation \cite{Baydin2018AutomaticSurvey} can help to calculate the derivatives accurately.  Merging two neural networks, the Heston-ANN and the IV-ANN, into a single network should make it more efficient when computing the implied volatility surface for the Heston model.


\bibliographystyle{abbrv}
\bibliography{mybib}

\end{document}